\documentclass[3p,preprint,sort&compress]{elsarticle}
\usepackage{amsmath}
\usepackage{amssymb}
\usepackage{amsfonts}
\usepackage{epsfig}
\usepackage{graphicx}
\usepackage{multirow}
\usepackage{caption}
\usepackage{cite}
\usepackage[usenames,dvipsnames]{xcolor}
\usepackage{tikz}
\usepackage{float}

\journal{ArXiv}

\begin{document}

\begin{frontmatter}

\title{Environmental Policy Regulation and Corporate Compliance\\ in a Spatial Evolutionary Game Model\\ }

\author[PUC]{Gabriel Meyer Salom\~ao}

\author[PUC]{Andr\'e Barreira da Silva Rocha\corref{cor1}}
\ead{andre-rocha@puc-rio.br}

\address[PUC]{Department of Industrial Engineering,
	Pontifical Catholic University of Rio de Janeiro,\\
	Rua Marqu\^es de S\~ao Vicente 225, G\'avea, CEP22451-900, Rio de Janeiro, RJ, Brazil.}

\cortext[cor1]{Corresponding author}

\begin{abstract}
We use an evolutionary game model to study the interplay between corporate environmental compliance and enforcement promoted by the policy maker in a country facing a pollution trap, i.e., a scenario in which the vast majority of firms do not internalize their pollution negative externality and auditors do not inspect firms. The game conflict is due to the trade-off in which firms are better-off when they pollute and are not inspected, while social welfare is maximized when auditors do not need to inspect socially responsible corporations that account for pollution in their production decisions regarding technology used and emission level. Starting with a well-mixed two-population game model, there is no long-run equilibrium and the shares of polluters and shirking auditors keep oscillating over time. In contrast, when firms and auditors are allocated in a spatial network, the game displays a rich dynamics depending on the inspecting cost. While the oscillatory behaviour is still possible, there is a set of parameters for which a long run robust equilibrium is achieved with the country leaving the pollution trap. On the other hand, an excessively high inspection cost leads to an ineffective auditing process where the few compliant firms are driven out of the country.  
\end{abstract}


\end{frontmatter}

\section{Introduction}
\label{sec:intro}
Environmental issues such as the level of emissions and global warming have become increasingly important during the past decades. 
On the one hand, there is a growing commitment among the countries' governments to global agreements such as the Kyoto Protocol
and the Paris Agreement. On the other hand, companies display significant efforts to address socially responsible issues raised by stakeholders such as consumers and employees. The latter is reflected in the accounting practices of the large companies, in which environmental and social accounting are frequently reported within the companies annual reports. Other important corporate strategies include the award of international certification stating compliance with social and environmental issues such as ISO 14001. 

The impact of social reputation and environmental compliance on profits is nowadays taken very seriously, particularly by large companies. Once a firm is inspected and declared non-compliant, there might be impact on revenues, market-share and costs: fines might be imposed by auditors and consumers can enforce product boycotts such as the one Royal Dutch Shell suffered due to the intention of the latter to dispose its Brent Spar offshore storage buoy in the sea (The Economist, 1995; Klein, Smith and John, 2002). Despite such a risk, social or environmental commitment and profits cannot be simultaneously maximized unless compliance investment contributes to profit maximization (Husted and Salazar, 2006). Consequently, environmental compliance has to be enforced by policy makers through auditing processes. But enforcement is also a costly process due to inspection costs, noisy monitoring due to the presence of multiple firms discharging pollutants in the same neighbourhood or even due to uninspectability (Heyes, 1998). The latter is particularly true in the US where the 4$^{th}$ Amendment obligates auditors to run the first inspections from outside the firms' fence. Corruption, privately funded political campaigns, bribery or the lack of auditors commitment, among other factors, may aggravate the situation.

Moreover, sanctions might depend on several factors such as the degree of non-compliance, i.e., the difference between the observed and the accepted pollution level, as well as on the effort demonstrated by the firm to minimize the social negative externality due to pollution. Firms showing an investment plan in clean production are rarely punished if found to be non-compliant with regulations (Arguedas and Hamoudi, 2004). As a result, firms tend to comply with environmental regulation only if the investment to do so is offset by the expected cost of being inspected and caught. The latter can be increased through different mechanisms such as higher fines imposed on non-compliant firms or decreasing the inspection costs, which might lead to a higher probability of inspection.

There is a wide literature on environmental compliance, particularly focusing on the conflict of interests between policy makers and corporations. One branch of the literature addressing such problem is based on the principal-agent framework. Arguedas and Hamoudi (2004) develop a principal-agent model in which sanctions depend on both the environmental technology employed by the firm as well as on the degree of non-compliance. The regulator chooses the pollution standard and the probability of inspection while the firm decides on the emission level and the environmental technology to be adopted. A cleaner technology leads to smaller fines when those are applied. Generally, principal-agent models are better suited when the aim is to optimize the structure of regulation and enforcement in a single-firm setting, as pointed out in Malik (2007). 

When dealing with more competitive markets, a game theoretical model should be used. The latter can be based either on classic or evolutionary game theory. Zhao et al. (2012) propose a game model to study the behaviour of manufacturers in response to drivers to reduce environmental risk and carbon emissions in the context of green supply chain management. In the same vein, Zhu and Dou (2007) propose an evolutionary game model to study the interplay between enterprises and the government during enforcement of regulation. While the companies face a cost of compliance, there are also benefits due to savings in energy consumption and lower discharge costs. Monitoring is costly to the government and the latter also faces pollution treatment costs when companies fail to implement green supply chain measures. Public revenues come from fines paid by non-compliant firms. 

In a different context, that of auditing processes dealing with financial reports, Anastasopoulos and Anastasopoulos (2012), propose an evolutionary game in which auditors might conduct either a basic or an extended (and more costly) set of audit procedures. Only the latter are able to distinguish between companies that commit or not intentional fraud in their reports. The authors point out that evolutionary game models have advantages when compared to their classic game theory counterparts. Evolutionary games are capable of explaining how players achieve the equilibrium, i.e., the game is dynamic and agents might review their actions over time, and if the stationary points of the game are stable or not. Moreover, evolutionary games allow for bounded rationality, which is closer to real world corporate decisions. As Arthur (1994) points out, the type of rationality assumed in economics demands much of human behaviour and breaks down under complicated problems. In such situations, psychologists tend to agree that humans think inductively with bounded rationality. Over time, feedback from the competitive environment comes in and individuals replace their strategies whenever better adapted actions display more successful results.

In light of all these facts, we present an evolutionary game to study a competitive market in which auditors might or not inspect firms. The latter might be environmentally compliant or not. We closely follow Arguedas and Hamoudi (2004) with regard to the profit functions of the firms depending positively on the emission level and on dirtier (i.e., cheaper) technologies. All firms discharge some level of pollution, thus generating a negative social externality. Fines increase the dirtier the technology a company chooses to use and on the level of emission. We assume an initial condition equivalent to a social context of pollution trap, i.e., where almost none of the firms are compliant and almost the totality of the auditors display a lenient culture of not inspecting the firms. 

We start with a base model where there is no spatial structure as in Zhu and Dou (2007) and Anastasopoulos and Anastasopoulos (2012), i.e., each auditor might inspect any firm in the population. Then we introduce one of the main contributions of our model, which is to assume that firms and auditors are spatially allocated at different geographical locations. The latter seems more realistic given that a firm tends to adopt similar practices as those of its local neighbours. Thus, a firm located in a region with several polluters would have incentives to do the same. On the side of auditors, it would be less likely that a particular auditor was selected to inspect a firm located far away due to cost constraints. While the framework adopting two well-mixed populations in which any firm can interact with any auditor recovers the same oscillatory pattern as in Anastasopoulos and Anastasopoulos (2012), without an evolutionary equilibrium in the long run, the extension assuming structured populations allows for a quite rich dynamics depending on the monitoring cost, with the possibility of different robust equilibria as well as the possibility of the oscillatory pattern observed in the well-mixed population case.  

Our structured population approach is closely related to cellular automata (CA) models. The latter have been used in a wide scope of problems, ranging from the seminal theoretical Game of Life by John Conway (see Gardner, 1970) to more applied problems such as the model in Duff et al. (2015) used to determine optimal travel routes for vehicles travelling from bases to forest fires. In CA models, an array of identically programmed cells (automata) interact with one another in a local neighbourhood. Each automaton has an initial state which might change or not over time according to some transition function. The latter specifies some rule(s) which generally depend on the state or some other entity of a particular cell and its closest neighbours. Local patters of behaviour then influence the entire population. Transition rules from the current to the next discrete time step might be completely deterministic as in Pereira et al. (2008) where the updating rule consists in every automaton copying the state of the best performing automata in the local neighbourhood or as in Conway's game of life where the state of each automaton could be alive or dead conditional on how many local neighbours were in the alive state in the previous time step. 

We here follow a probabilistic transition rule in which each cell contains one auditor and one firm. Individuals compare their performance within their own population although their profits come from the game played with individuals in the other population. Updating is synchronous such that, in each Montecarlo step (MSC), after all individuals have computed the payoffs obtained from the game, each individual selects a local neighbour to compare payoffs. If the neighbour's payoff is larger, there is an increasing probability of copying the neighbour's strategy the larger the difference in payoffs is. Games where players compare their performance and update their strategies probabilistically can be found in the literature as in Chen et al. (2015), although generally such class of models only deals with one-population games.

The remainder of the paper is organized as follows. In Section 2, we setup and solve the evolutionary game model assuming well-mixed populations. In Section 3, the game is extended to the case of structured populations. Results are compared and discussed. Section 4 concludes.  

\section{Model}
\label{sec:model}
We follow Arguedas and Hamoudi (2004) with regard to firm profit and social welfare function. A firm's private profit $\pi(k,e,\beta)=ke-e^2\beta^{-1}$ is strictly concave on the emission level $e>0$ and is monotonically increasing on a profitability parameter $k>0$ and on the type of production technology employed $\beta\in[1,\bar{\beta}]$, where a higher $\beta$ is associated with a cheaper and dirtier technology. Pollution generates a negative externality $d(e,\beta)=\beta e^2$ which is convex increasing on the emission level and linearly increasing the dirtier the technology adopted. On the policy maker side, auditors might inspect firms or not. If a firm is inspected and declared non-compliant, it faces a fine $f(e,\beta,s)=\beta(e-s)^2$, where $s$ is the maximum emission level acceptable. 

Differently from the principal-agent monopoly framework in Arguedas and Hamoudi (2004), we assume a very large number of firms in a competitive market, thus making inspectability through emission level more difficult (see Heyes, 1998). We thus normalize the emission standard to $s=0$, making fines harsher, i.e., due to a higher required effort to correctly measure the emission level, the policy maker adopts a stricter environmental policy. The decision to punish is solely based on visual inspection of the type of technology employed in corporate plants and punishment is enforced whenever $\beta$ is high. The penalty consists in charging a fine equal to $d(e,\beta)$, i.e., a fine equivalent to the complete internalization of the negative environmental externality that a firm generates while producing. 

We also assume bounded rational agents such that the decision to select the optimal technology is carried out through maximizing profits but the profit maximizing function that is chosen by managers is purely based on their corporate culture. In other words, when firms decide which technology to use, environmentally compliant firms internalize the negative externality in their private profit function while non-compliant firms do not, i.e., the choice of environmental policy is naive and agents do not know in the long run which corporate policy is the best performing strategy. Such an approach is similar to that used in Xiao and Yu (2006) supply chain model, where agents rationally maximize their utilities (either revenue or profit functions) but they are naive regarding which preference function is the best over time. 
 
Regarding inspection, there is a very large number of auditors, which are randomly allocated to visit firms. Each auditor either inspects or not. While in Arguedas and Hamoudi (2004) the inspection probability faced by the only firm in the model was a continuous variable $p\in[0,1]$, here it can only assume one of two values for one particular firm at the microscopic local level, i.e., $p=0\vee p=1$, depending on the strategy chosen by the auditor visiting the firm. But at the macroscopic level, every firm on average faces $p\in[0,1]$ depending on the strategy distribution (state) in the auditors population at a given time step. With regard to welfare, the policy maker has a social welfare function according to which he takes into account the firm's private profit, the negative externality it generates and the expected cost of inspection. Fines are not taken into account once the fine paid by a firm cancels out with the public revenue it generates:
\begin{equation}
W(k,e,\beta,p,c)=ke-e^2\beta^{-1}-\beta e^2-pc
\label{welfare}
\end{equation}
Whenever an auditor visits a firm, the game played is given by:
\begin{equation}
\bordermatrix {&\text{Non-compliant (P)}&\text{Compliant (C)}\cr
	\text{Inspects (I)} &\pi_A^{IP},\pi_F^{IP}& \pi_A^{IC},\pi_F^{IC} \cr
	\text{No Inspection (N)} &\pi_A^{NP},\pi_F^{NP}& \pi_A^{NC},\pi_F^{NC}\cr} 
\label{Matrix-I}
\end{equation}
where the auditor, denoted by subscript $A$ (resp. the firm, denoted by subscript $F$), stands as the row (resp. column) player in the payoff matrix above. When a firm is called to play the stage-game in (\ref{Matrix-I}), all the managers can do is to adjust the emission level selecting the optimal emission that maximizes the profit function depending if the auditor the firm faces will inspect it or not and depending on which environmental policy the firm follows. In other words, we further assume that the optimal decisions regarding $\beta$ and $e$ are taken rationally but independently at each firm. This assumption, which is also in line with bounded rationality, takes into account the lack of synergy between corporate divisions where the long-run investment decision regarding the technology to be used is made by a planning team while the more short-run decision on the emission level is carried out by the operating team. 

If the firm is compliant with the environmental regulation, technology choice is based on the following maximization problem:
\begin{equation*}
\max_{\beta}\pi(k,e,\beta)=ke-e^2\beta^{-1}- \beta e^2
\end{equation*}
which internalizes the negative externality into the firm's private profit for the sake of selecting $\beta^*$. First order condition (FOC) leads to:
\begin{equation*}
\frac{\partial\pi}{\partial\beta}=-e^2\left( 1-\beta^{-2}\right)=0\Rightarrow\beta^*=1
\end{equation*}
Instead, a non-compliant firm only takes into account its private profit function, leading to the following FOC and technology optimal choice:
\begin{equation*}
\frac{\partial\pi}{\partial\beta}=e^2\beta^{-2}>0\Rightarrow\beta^*=\bar{\beta}
\end{equation*}
Given that we are left with only two choices of technology $\beta\in\left\lbrace1,\bar{\beta} \right\rbrace$, whenever an auditor visits a firm and decides to inspect it, the firm is fined if it employs a technology $\beta=\bar{\beta}$. The optimal emission level for a compliant firm is independent on being audited or not:
\begin{equation*}
\max_{e}\pi(k,e,1)=ke-e^2-e^2\Rightarrow\frac{\partial\pi}{\partial e}=k-4e=0\Rightarrow e^*=k/4
\end{equation*}
leading to a private profit $\pi_F^{IC}=\pi_F^{NC}=3k^2/16$ and social welfare $\pi_A^{NC}=k^2/8$ if the firm is not inspected and $\pi_A^{IC}=k^2/8-c$ otherwise. We emphasize that, although compliant firms always internalize the pollution cost into their private profit function for the sake of selecting their optimal decisions $\beta^*$ and $e^*$, they never get fined, thus never facing the social cost $d(e,\beta)$ due to pollution.   
Regarding non-compliant firms, if a firm is not inspected, it is not declared non-compliant and does not have to internalize the environmental damage into its private profit function. The FOC for profit maximization leads to:
\begin{equation*}
\max_{e}\pi(k,e,\bar{\beta})=ke-e^2\bar{\beta}^{-1}\Rightarrow\frac{\partial\pi}{\partial e}=k-2e\bar{\beta}^{-1}=0\Rightarrow e^*=k\bar{\beta}/2
\end{equation*}
Thus, the firm optimal private profit is $\pi_F^{NP}=k^2\bar{\beta}/4$, leading to a corresponding social welfare of $\pi_A^{NP}=k^2\bar{\beta}(1-\bar{\beta}^2)/4$. On the other hand, a non-compliant firm facing inspection is fined and has to internalize the pollution cost, not only when making the optimal emission decision but also decreasing its private profit that is left to be reinvested. The optimization problem becomes:
\begin{equation*}
\max_{e}\pi(k,e,\bar{\beta})=ke-e^2\bar{\beta}^{-1}-\bar{\beta} e^2
\end{equation*}
leading to $e^*=k\bar{\beta}(1+\bar{\beta}^2)^{-1}/2$, $\pi_F^{IP}=k^2\bar{\beta}(1+\bar{\beta}^2)^{-1}/4$ and $\pi_A^{IP}=k^2\bar{\beta}(1+\bar{\beta}^2)^{-1}/4-c$. 
From the analysis above, while a compliant firm always voluntarily internalizes the pollution damage into its profit function whenever managers make optimal decisions about technology and emission level, non-compliant firms only do so when an auditor inspects its non-switchable production technology and imposes environmental compliance through the fine. The latter has the effect of adjusting the firm's pollution discharge to a lower emission level. While a non-compliant uninspected firm is better-off than any other firm with regard to private profit, i.e., $k^2\bar{\beta}/4>3k^2/16>k^2\bar{\beta}(1+\bar{\beta}^2)^{-1}/4$, a non-compliant inspected firm faces the lowest possible payoff. Despite employing the cheapest technology, getting punished and having to adjust the emission level using the worst technology leads to a too low optimal emission level, i.e.,  $k\bar{\beta}(1+\bar{\beta}^2)^{-1}/2<k/4<k\bar{\beta}/2;\ \forall\bar{\beta}>1$. On top of this, the firm's private profit is also decreased due to the additional cost to pay the fine.

The payoff matrix in (\ref{Matrix-I}) with possible private profits $\pi_F$ and social welfares $\pi_A$ then becomes:
\begin{equation}
\bordermatrix {&\text{Non-compliant (P)}&\text{Compliant (C)}\cr
	\text{Inspects (I)} &k^2\bar{\beta}(1+\bar{\beta}^2)^{-1}/4-c,\ k^2\bar{\beta}(1+\bar{\beta}^2)^{-1}/4& k^2/8-c,\ 3k^2/16 \cr
	\text{No Inspection (N)} &k^2\bar{\beta}(1-\bar{\beta}^2)/4,\ k^2\bar{\beta}/4& k^2/8,\ 3k^2/16\cr} 
\label{Matrix-II}
\end{equation}
From (\ref{Matrix-II}), enforcement of environmental compliance is only feasible if $k^2\bar{\beta}(1+\bar{\beta}^2)^{-1}/4-c>k^2\bar{\beta}(1-\bar{\beta}^2)/4$, leading to a maximum inspection cost $c<\bar{c}=k^2\bar{\beta}^5(1+\bar{\beta}^2)^{-1}/4$. Above $\bar{c}$, not inspecting is a strictly dominant strategy for auditors, independently of the corporate culture the visited firm follows, and society falls into a pollution trap scenario in the long run where polluters always perform better than any compliant firm, thus driving the latter to extinction. Moreover, the largest social welfare $k^2/8$ is achieved whenever a compliant firm does not require any inspection effort. In contrast, the largest private profit that can be achieved $k^2\bar{\beta}/4$ is obtained when auditors fail to inspect a non-compliant firm. Hence, the game payoff matrix points out the conflict of interest between the policy maker and firms in the competitive market.

\subsection{Mean field approximation: solving the well-mixed population game}
\label{sec:game}
In this section, we start analysing the evolution of the states of both populations playing the game in (\ref{Matrix-II}) using replicator dynamics (RD). The latter is suitable when populations are well-mixed, i.e., any firm can be visited by any auditor. Moreover, RD assumes replication, i.e., any profit obtained by a compliant (resp. non-compliant) firm is completely reinvested in another compliant (resp. non-compliant) firm. Under RD only natural selection of strategies is taken into account and there is no mutation, i.e., a firm does not reinvest in another firm experimenting a different corporate culture. Assuming that at a given time step $p$ is the proportion of auditors that inspect firms and $q$ is the proportion of non-compliant firms, one can derive a system of nonlinear ordinary differential equations governing the evolution of both populations over time. Each equation of the RD system governs the change in the state of one population over time. For the case of our game, we have:
\begin{equation}
\frac{\partial p}{\partial t}=\dot{p}=p(1-p)(\pi_A^{I}-\pi_A^{N})\Rightarrow \dot{p}=p(1-p)\left(\frac{k^2\bar{\beta}^5}{4(1+\bar{\beta}^2)}q-c \right)
\label{RDp}
\end{equation}
\begin{equation}
\frac{\partial q}{\partial t}=\dot{q}=q(1-q)(\pi_F^{P}-\pi_F^{C})\Rightarrow \dot{q}=q(1-q)\left(\frac{k^2(4\bar{\beta}-3)}{16}-\frac{k^2\bar{\beta}^3}{4(1+\bar{\beta}^2)}p \right)
\label{RDq}
\end{equation}
where $\pi_A^{I}$ and $\pi_A^{N}$ (resp. $\pi_F^{P}$ and $\pi_F^{C}$) are respectively the expected social welfares when an auditor inspects and does not inspect a firm (resp. the expected private profits for a non-compliant and a compliant firm). The RD system in (\ref{RDp}-\ref{RDq}) does not have an evolutionary equilibrium (Hofbauer and Sigmund, 1998). It has a neutrally stable state $p^*=\bar{\beta}^{-3}(4\bar{\beta}-3)(1+\bar{\beta}^2)/4;\ q^*=4c(1+\bar{\beta}^2)k^{-2}\bar{\beta}^{-5}=c/\bar{c}$, corresponding to the only (mixed strategy) Nash equilibrium of the stage-game in (\ref{Matrix-II}), about which the states of both populations oscillate over time indefinitely. Such evolutionary pattern is similar to that found in Anastasopoulos and Anastasopoulos (2012). 

\begin{figure}[htp]
	\centering
	\begin{tabular}{cc}
		\epsfig{file=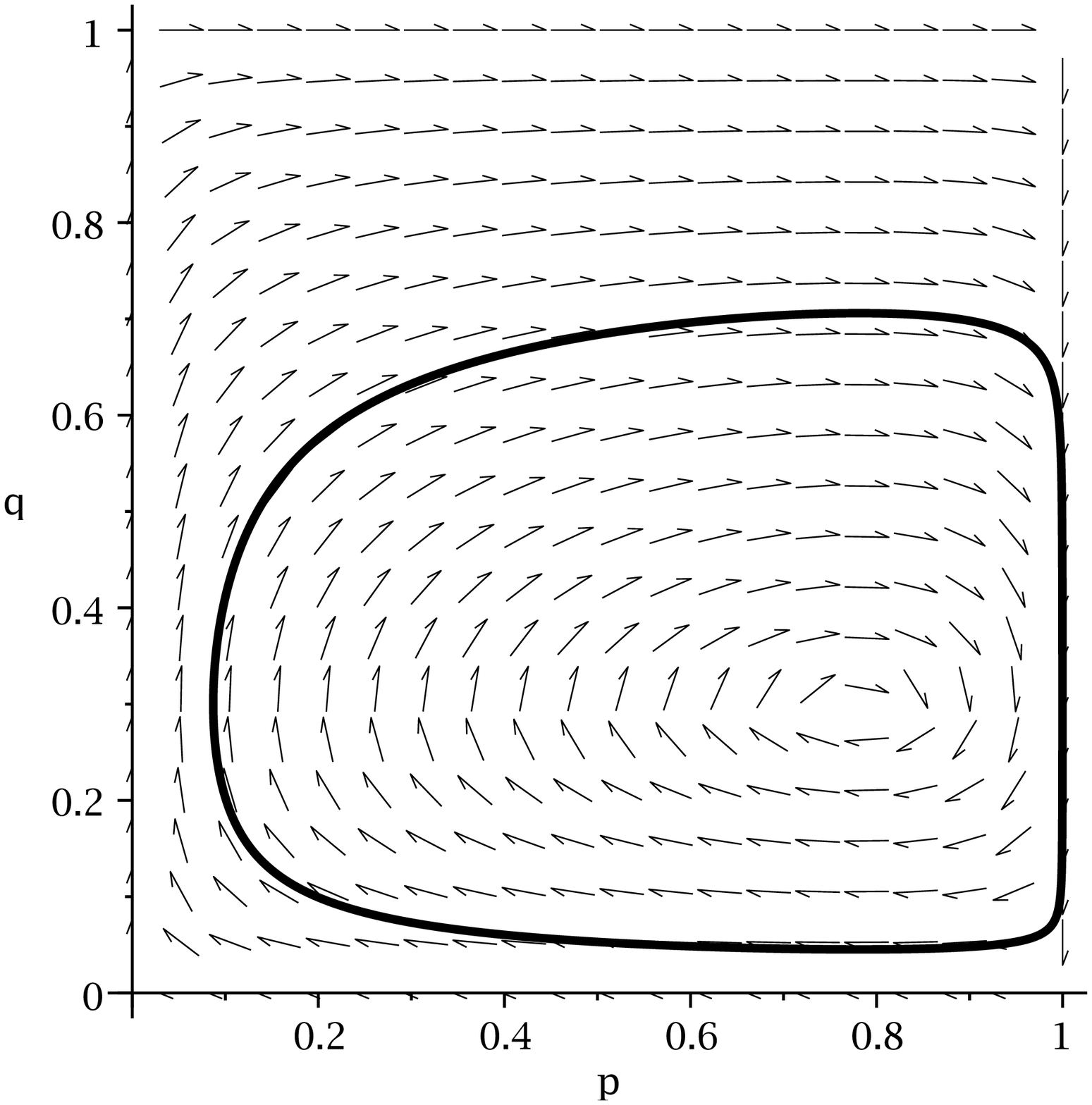,height=5cm,width=5cm,angle=0}&
		\epsfig{file=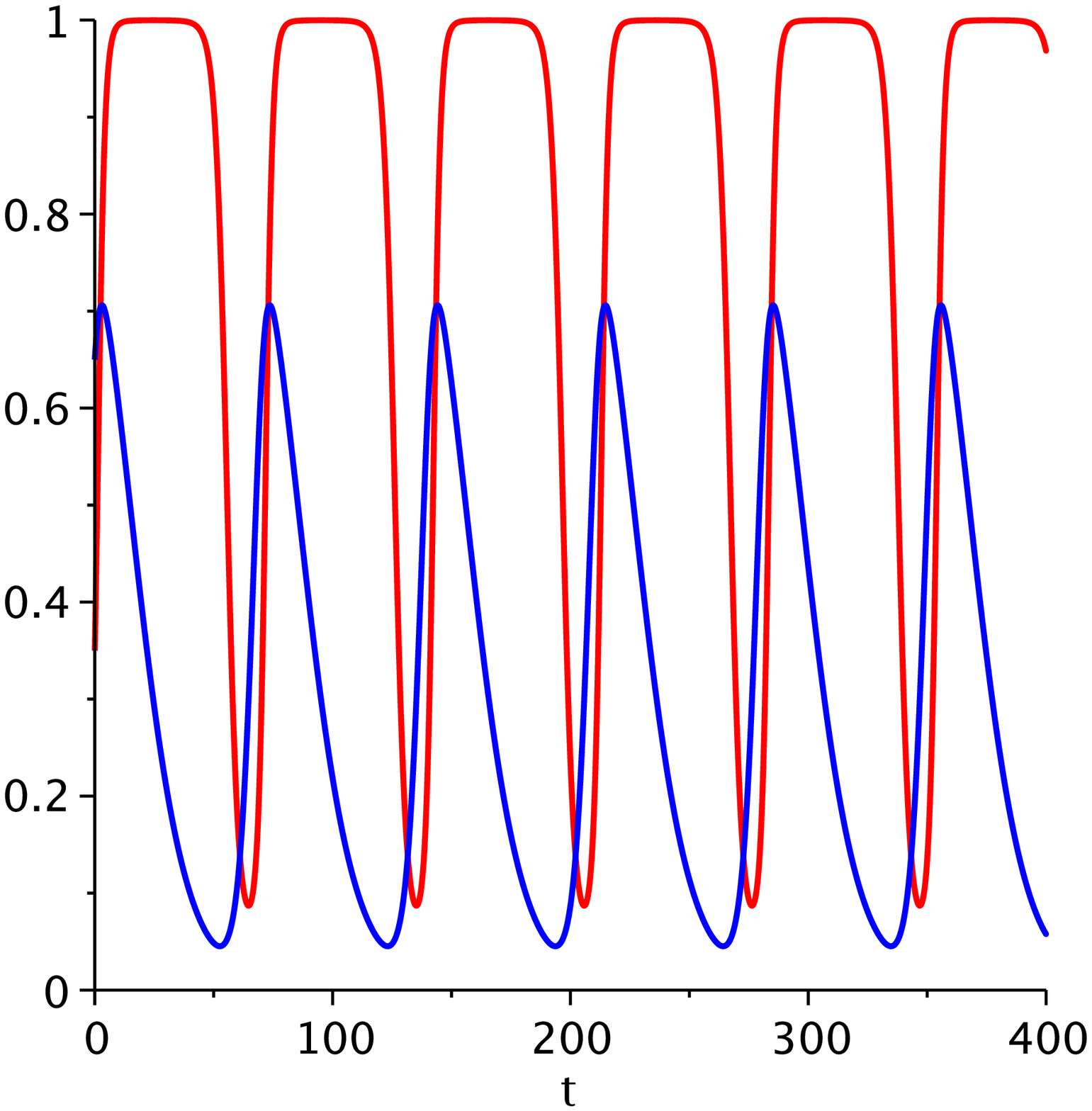,height=5cm,width=8cm,angle=0}
	\end{tabular}
	\vspace{.5cm}
	\caption{(color online) Phase diagram and time evolution for parameters: $k=1$, $\bar{\beta}=2$, $c=0.30\bar{c}$ and initial conditions $(p_0=0.35,q_0=0.65)$.}
	\label{fig:1}
\end{figure}

In Figure \ref{fig:1} we present a numerical simulation for $k=1$, $\bar{\beta}=2$, $c=0.30\bar{c}=0.48$ and initial conditions in which only 35\% of the firms are compliant and only 35\% of the auditors are committed to inspect firms, $p_0=0.35\wedge q_0=0.65$. It can be seen that compliance and inspection keep oscillating in both populations without achieving an equilibrium. The initial proportion of two-thirds of the firms being non-compliant induces a sharp rise in the proportion of inspecting auditors over time. The increase in the likelihood of being inspected favours the increase of compliance among firms. When the proportion of non-compliant firms has decreased sharply, close to 5\% of the population of firms, auditors become more lenient and the likelihood of inspecting falls down to 10\%. The cycle then restarts.    

Regarding the neutrally stable state $(p^*,q^*)$, it represents the average behaviour of the populations over a long interval of time (Cressman et al., 1998). Given that $\lim_{\bar{\beta}\rightarrow 1} p^*=0.5$, one can see that there is a bias towards inspection, i.e., on average the majority of the auditors inspects the firms they visit. Moreover, from (\ref{Matrix-II}), the expected profit of a non-compliant firm is given by $\pi_F^{P}=\frac{k^2\bar{\beta}}{4(1+\bar{\beta}^2)}\left[ 1+(1-p)\bar{\beta}^2\right] $, thus $\frac{\partial\pi_F^{P}}{\partial p}=-\frac{k^2\bar{\beta}^3}{4(1+\bar{\beta}^2)}<0$, i.e., given $\bar{\beta}\wedge k$, whenever the probability of inspection increases, the expected profit of non-compliant firms decreases. Consequently, given the expected profit of compliant firms $\pi_F^{C}=\frac{3k^2}{16}$ is unaffected by changes in $p$, the incentive to comply with the environmental regulation increases. On the side of the policy maker, the difference between the expected social welfare with and without inspection is given by $\pi_A^{I}-\pi_A^{N}=\frac{k^2\bar{\beta}^5}{4(1+\bar{\beta}^2)}q-c$. Thus, $\frac{\partial\left(\pi_A^{I}-\pi_A^{N} \right) }{\partial q}>0$ and $\frac{\partial\left(\pi_A^{I}-\pi_A^{N} \right) }{\partial c}<0$. Hence, given $\bar{\beta}\wedge k$, whenever non-compliance spreads in the population of firms, the incentive for inspection increases as well. On the other hand, larger costs of inspection induce a lower incentive to inspect firms. This analysis is valid $\forall\bar{\beta};\ \forall k$.  

One drawback of employing RD to model the evolution of both populations over time is that, by assuming well-mixed populations, RD does not take into account the geographical location of firms and auditors. It is equivalent to a network where all agents are neighbours of each other. In the next section, we overcome this issue by assuming that both populations are spatially located in an $L\times L$ square grid with periodic boundary conditions. In such setup, firms and auditors only interact with their local geographic neighbours.   

\section{Spatial Game}
\label{sec:spatial}
In order to increase realism, we extend the game in Section \ref{sec:model} assuming individuals in both populations keep playing the same stage-game in (\ref{Matrix-II}) but now each player has a specific geographical location. While it seems obvious to account for firms' spatial location, it is also reasonable to make the same assumption with regard to auditors. Generally, public servants are allocated to work in a particular region of the country or state and they are in charge of inspections in the neighbourhood of the area where they work. 

The extended game model considers a regular square grid of size $L\times L$, with periodic boundary conditions, thus a planar representation of a torus. We assumed each population with $L^2=10,000$ individuals such that there is no noise due to small population effects. Each location of the grid is occupied by two players, one auditor and one firm, and each player has a von-Neumann (vNM) neighbourhood with four neighbours. At the beginning of each simulation, each firm in the grid is randomly allocated either strategy $C$ or $P$ according to some pre-defined initial condition $q_0$. Regarding the population of auditors, we assumed the government selects a particular region of the country to start a pilot task force to try to promote inspection within the entire population of auditors. We focused on countries starting from a pollution trap scenario in which $p_0=0.05\wedge q_0=0.95$. Thus, at the initial conditions, all auditors are allocated strategy $N$, except in a strip representing a compact spatial cluster composed of $p_0 L$ auditors playing $I$.

Once the allocation is carried out, every individual $i$ in both populations plays the game against all individuals in the opponent population allocated in his vNM neighbourhood plus the opponent allocated at the same location he occupies. The sum of the payoffs obtained in each contest is computed, $\Pi_i=\sum_{i=1}^{n=5}\pi_i$, and a random neighbour $j$ from the same population is chosen to compare payoffs. If $\Pi_i>\Pi_j$, player $i$ keeps his strategy, otherwise his strategy is updated with probability $(\Pi_j-\Pi_i)/\left[ 5(\max\pi_i-\min\pi_i)\right] $, where $\max\pi_i$ (resp. $\min\pi_i$) is the maximum (resp. minimum) attainable payoff player $i$ can get in the stage-game payoff matrix. After all individuals have had the chance to update their strategies, the MCS ends and both populations are updated synchronously. 

We carried out numerical simulations keeping $k=1$ and $\bar{\beta}=2$ as in Section \ref{sec:model} and varied parameter $\xi=c/\bar{c}\in[0.05;0.99]$ given that, as seen in subsection \ref{sec:game}, the inspection cost plays an important role by decreasing the incentive to inspect and, consequently, decreasing the firms' long run average compliance behaviour. For all values of $\xi$, the stage-game still has one unique mixed-strategy Nash equilibrium, leading to the same cyclical evolutionary pattern found in subsection \ref{sec:game} when both populations are well-mixed. We found that, despite this fact, the numerical results displayed a quite rich dynamics in the spatial game. Moreover, in more than one case, a robust long-run equilibrium is attainable in both populations.

We used $\xi=\left\lbrace 0.05;0.25;0.40;0.50;0.55;0.60;0.65;0.75;0.90;0.99\right\rbrace$. We ran a set of 50 simulations for each value of $\xi$, with each realization using a different seed. Each realization ran for 5,000 MCS, which was more than enough to ensure either convergence towards the long run equilibrium or to display a clear cyclical long run behaviour. The only exception was for $\xi=0.99$, which never displayed a cyclical behaviour but long run equilibrium required always a larger number of MCS to be achieved. Particular interest was given to the evolution of $p$ and $q$ over time, the snapshots of strategy distribution in the grid for both populations and the long run average state of each population when cyclical behaviour was displayed. Based on the numerical results, we formulate three propositions.\\

\textbf{Proposition 1:} For low to intermediate inspection cost, there is a long run robust evolutionary equilibrium in which the pollution trap scenario disappears. All auditors inspect firms and the latter comply with environmental regulations.

\begin{figure}[htp]
	\centering
	\begin{tabular}{cc}
		\epsfig{file=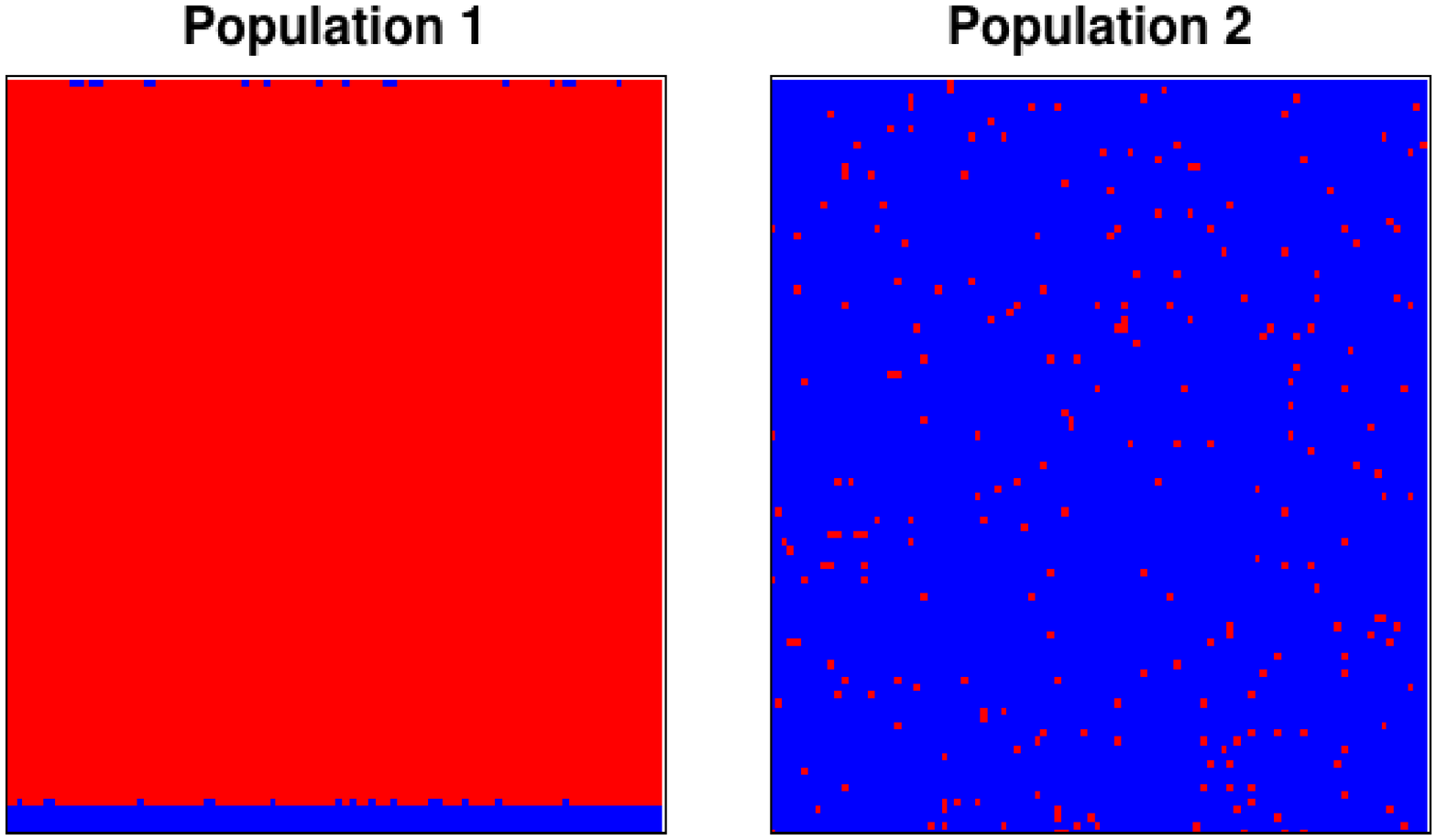,height=3.95cm,width=8cm,angle=0}&
		\epsfig{file=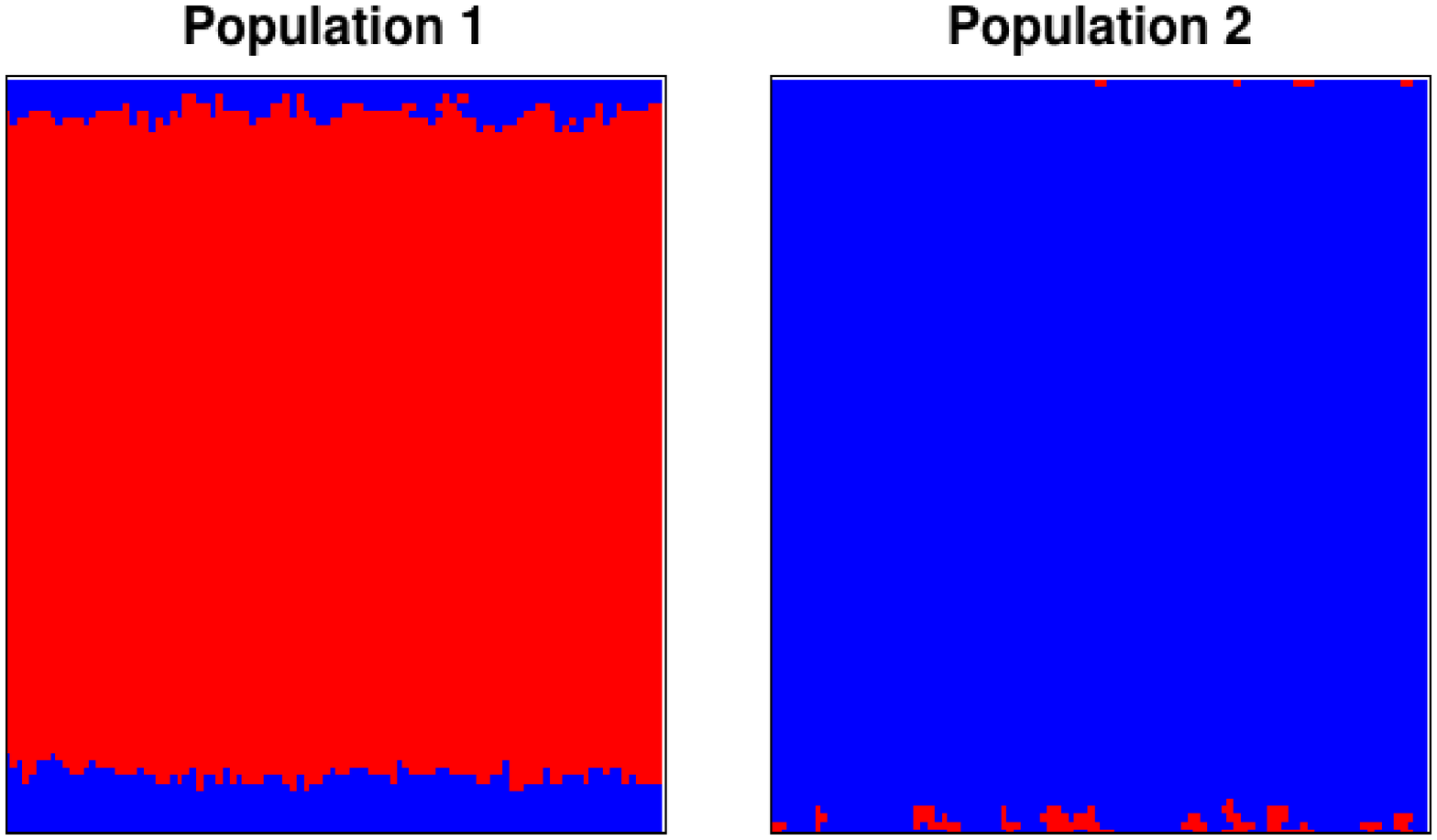,height=3.95cm,width=8cm,angle=0}\\
		\epsfig{file=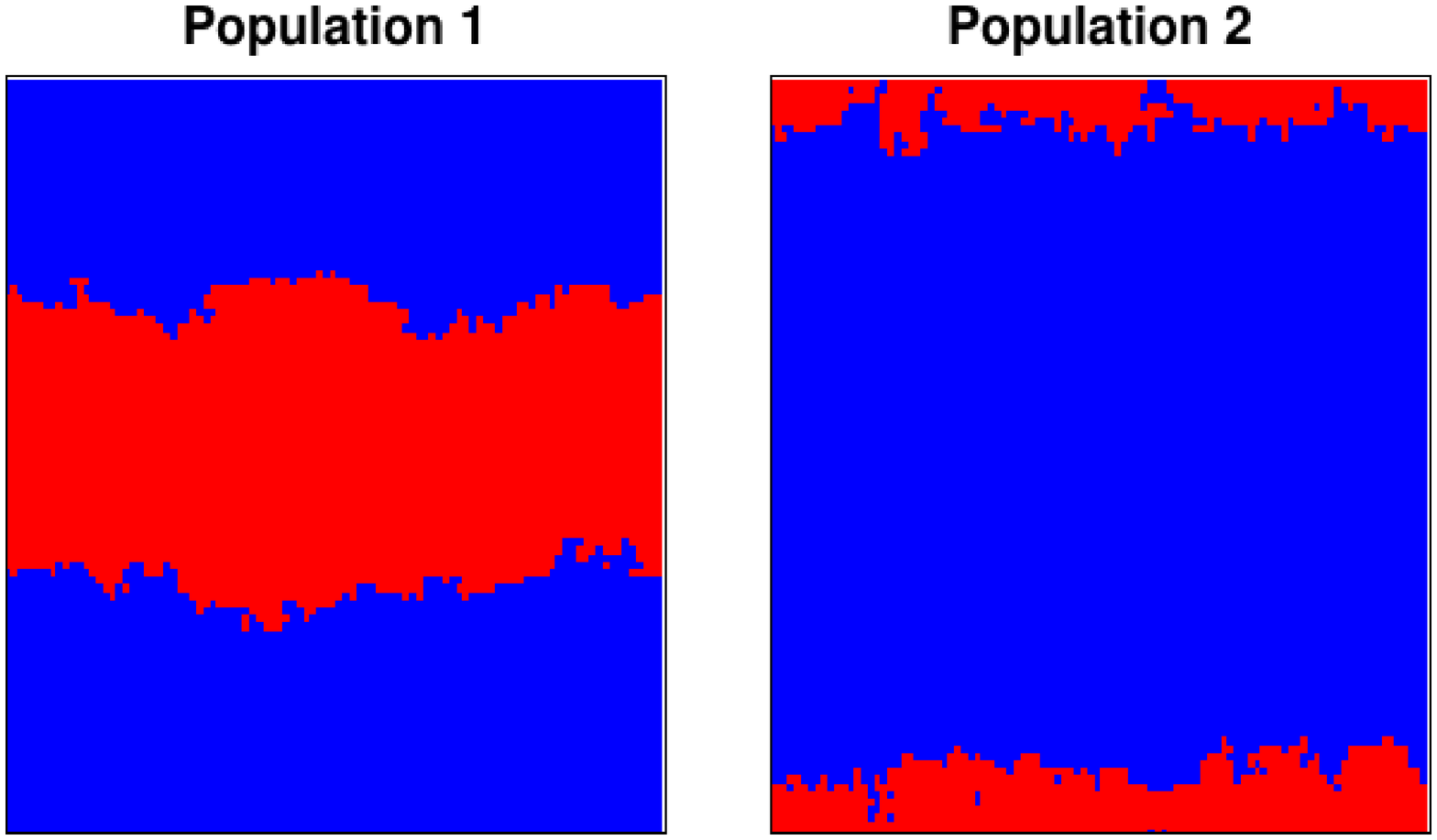,height=3.95cm,width=8cm,angle=0}&
		\epsfig{file=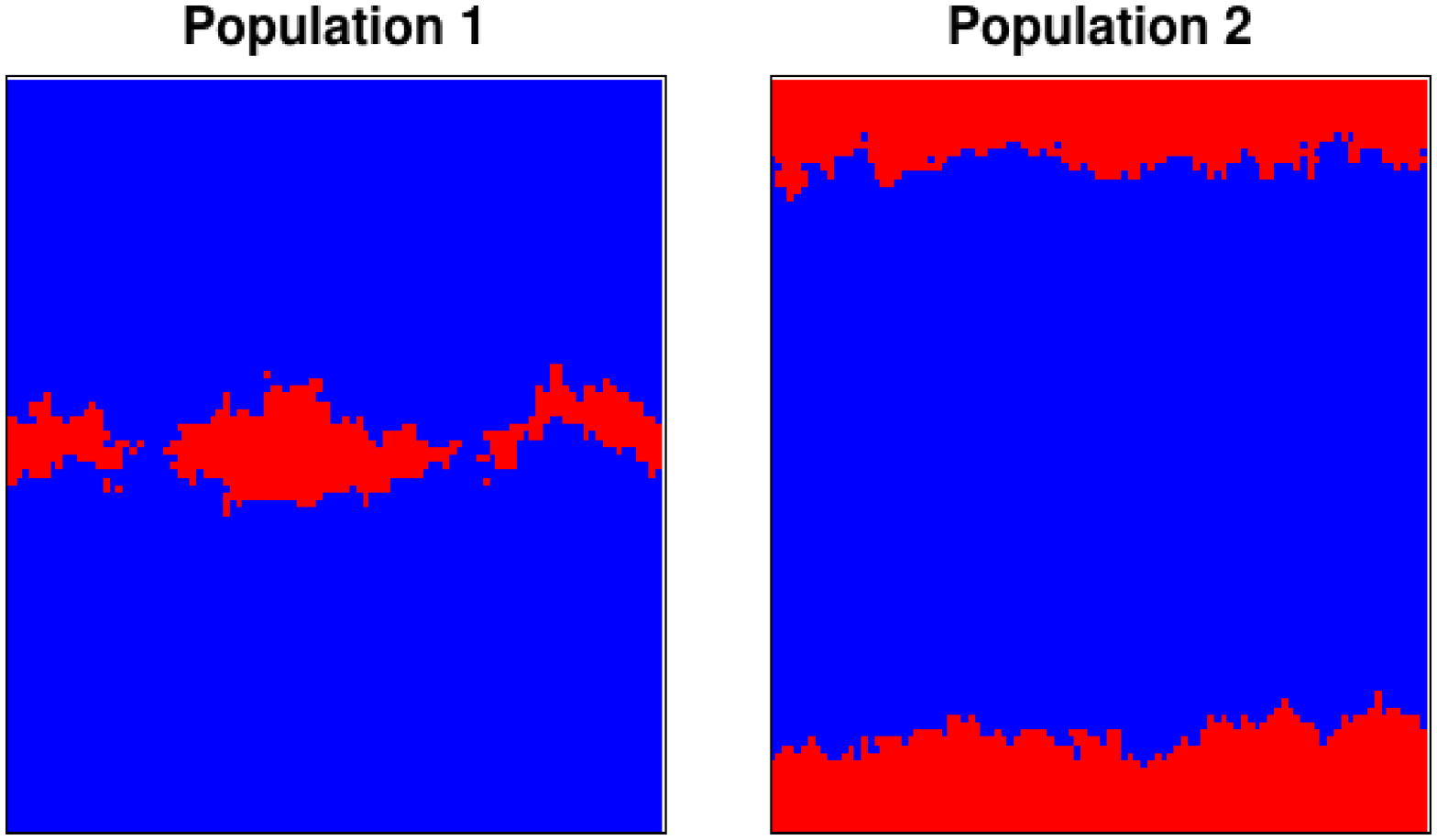,height=3.95cm,width=8cm,angle=0}\\
		\epsfig{file=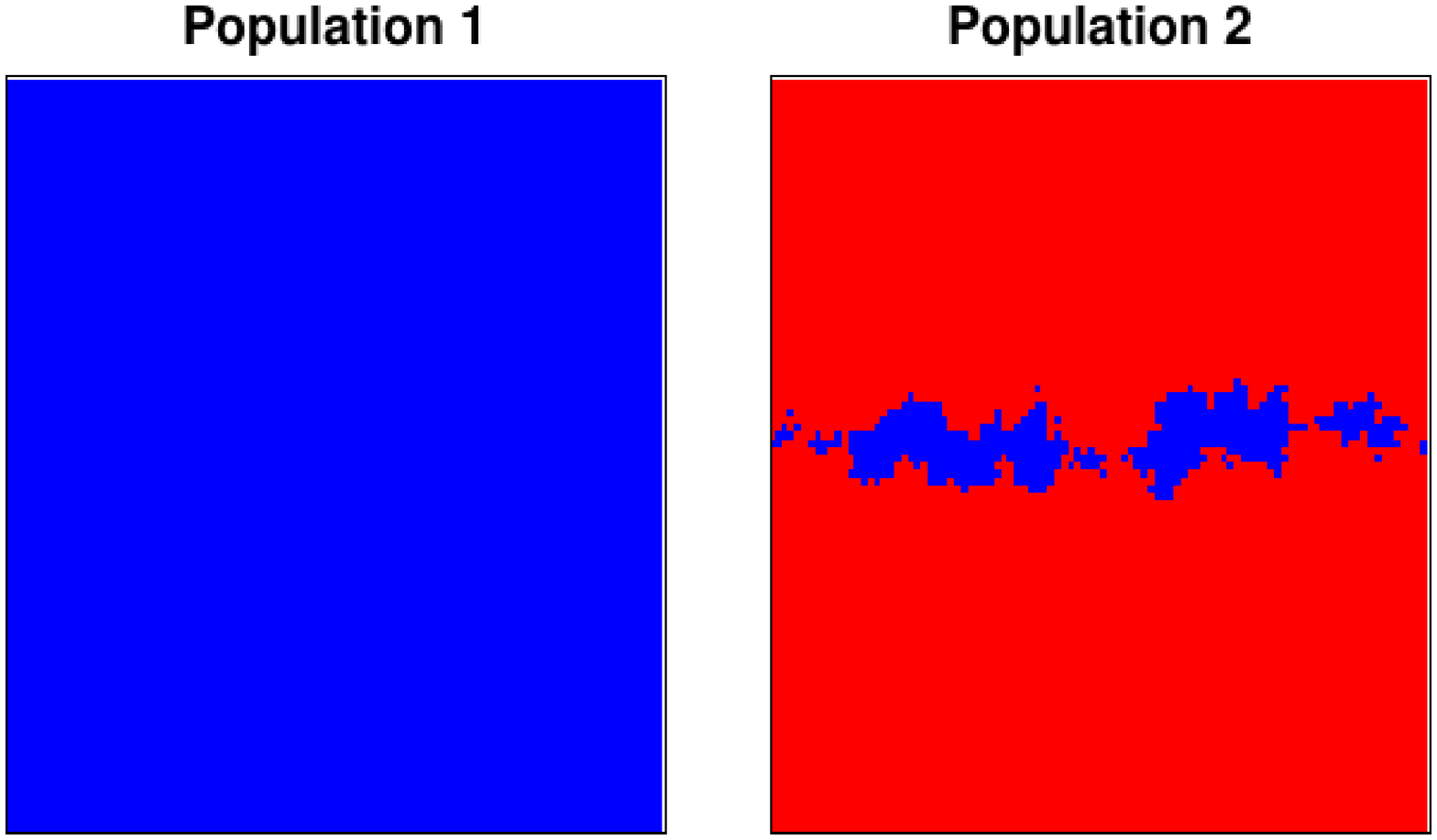,height=3.95cm,width=8cm,angle=0}&
		\epsfig{file=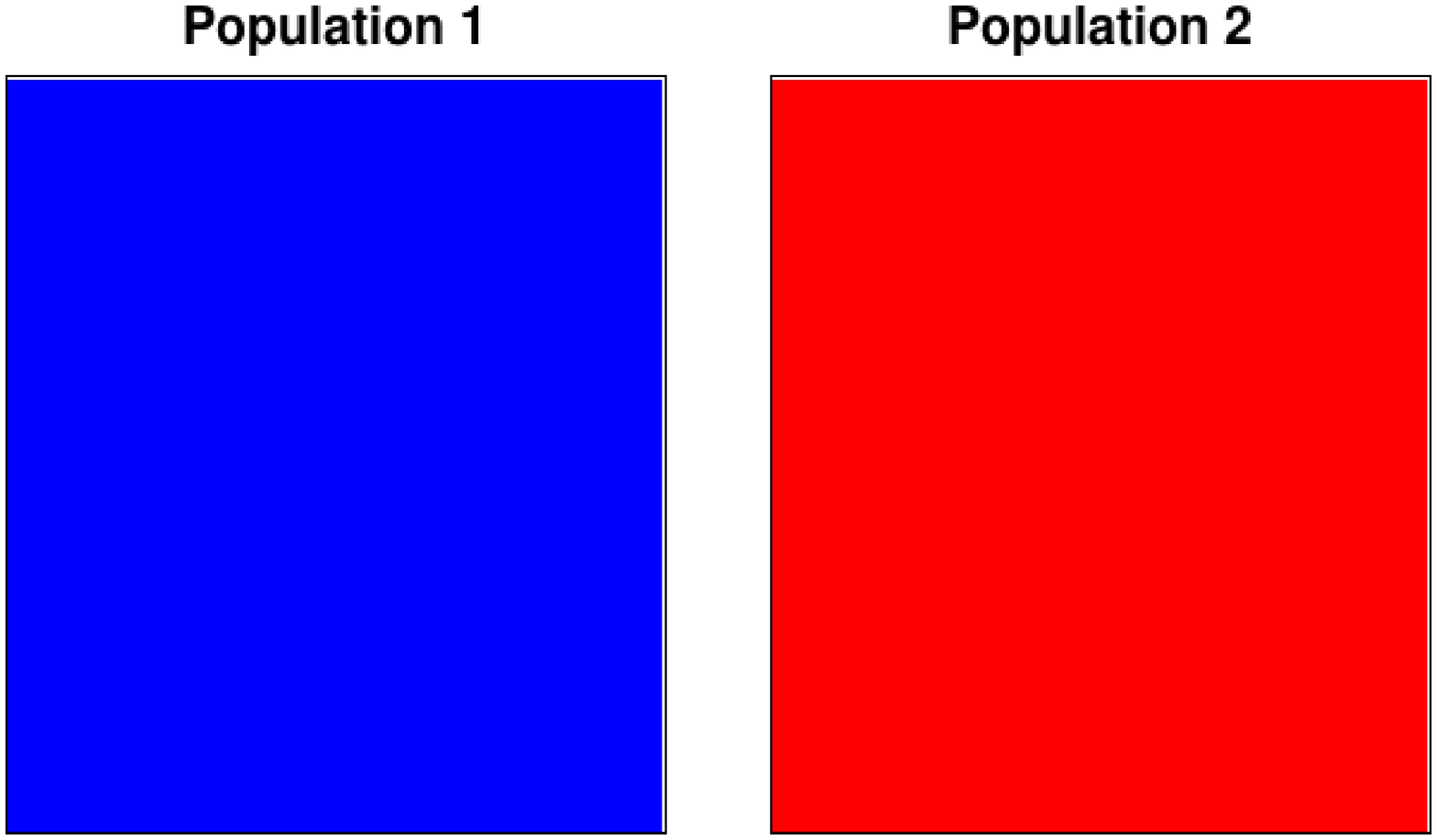,height=3.95cm,width=8cm,angle=0}
	\end{tabular}
	\caption{(color online) evolution of the state of each population for parameters: $k=1$, $\bar{\beta}=2$, $\xi=0.25$ and initial conditions $(p_0=0.05,q_0=0.95)$. In both populations, blue (resp. red) corresponds to strategy 1 (resp. 2). From top left to bottom right, square lattice at $T=1;17;100;150;500;800$.}
	\label{fig:2}
\end{figure}

\begin{figure}[htp]
	\centering
	\begin{tabular}{c}
		\epsfig{file=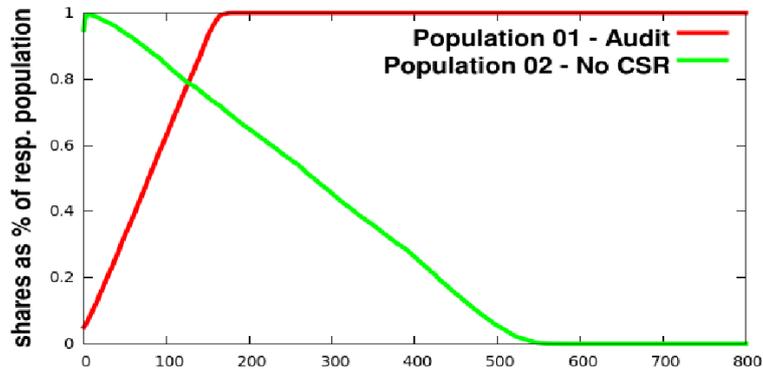,height=5cm,width=10cm,angle=0}
	\end{tabular}
	\caption{(color online) time evolution of the state of each population for $k=1$, $\bar{\beta}=2$, $\xi=0.25$ and $T=\left[ 0,800\right] $ MCS.}
	\label{fig:3}
\end{figure}

\textbf{Discussion:}
Using the case when $\xi=0.25$, a low cost of inspection, the corresponding payoff matrix in (\ref{Matrix-II}) becomes:
\begin{equation}
\bordermatrix {&\text{Non-compliant (P)}&\text{Compliant (C)}\cr
	\text{Inspects (I)} &-0.300,0.100& -0.275,0.1875 \cr
	\text{No Inspection (N)} &-1.500,0.500 & 0.125,0.1875\cr} 
\end{equation}

with just one mixed-strategy Nash equilibrium. In the well-mixed population model, as shown in subsection \ref{sec:game}, this corresponds to a dynamics in which the state of each population would oscillate over time without achieving a long run equilibrium. In contrast to the well-mixed population game, for all simulations performed with $\xi\in\left[  0.05;0.55\right] $, non-compliant firms and no-inspecting auditors were led to extinction in the long run. In Figure \ref{fig:2} we present the $L\times L$ square lattice for a particular realization using $k=1$, $\bar{\beta}=2$, $\xi=0.25$ and initial conditions $p_0=0.05\wedge q_0=0.95$ (pollution trap). The figure displays the evolution of the strategy distribution in each population from Montecarlo step $T=1$ to $T=800$. In Figure \ref{fig:3} the evolution of both populations over time is displayed. 

At $T=1$, one can see the thin blue strip with $\approx p_0L=500$ inspecting auditors in population 1 and the red dots corresponding to the $\approx(1-q_0)L=500$ compliant firms geographically spread in population 2. Due to a state in which 95\% of firms are non-compliant in population 2, the best response for auditors is to inspect ($\pi_A^{IP}=-0.300>\pi_A^{NP}=-1.500$). But the vast majority of the non-inspecting auditors are neighbours to each other, hence, even if they perform worse than inspecting auditors in terms of payoff, they do not switch strategy. The exception to the latter occurs at the boundaries between both clusters of inspecting and non-inspecting auditors. At the boundaries, inspecting auditors perform better and non-inspecting auditors start to switch strategy. The cluster with inspecting auditors remains compact and starts to get larger ($T=17$), covering already more than half of the grid area by $T=100$, until non-inspecting auditors are driven out from population 1.

In contrast, in population 2, compliant firms have already disappeared by $T=17$, except in the region corresponding to the cluster of inspecting auditors in population 1 (Figure \ref{fig:3} shows that during the initial MCS non-compliance actually increases in population 2). In that region, compliant firms perform better ($\pi_F^{IP}=0.100<\pi_F^{IC}=0.1875$) and very small compact clusters of compliant firms start to form. As the cluster of inspecting auditors becomes larger in population 1, the clusters of compliant firms start to merge ($T=100$), becoming a single cluster that keeps growing ($T=150$) until driving all non-compliant firms out from population 2. 

Although the evolutionary process seems straightforward, there is one key element for the success of this long run equilibrium ($p^*=1,q^*=0$): the speed at which the cluster of inspecting auditors spreads in population 1 is much faster than the spreading speed of the cluster of compliant firms in population 2. And this fact remains true until only inspecting auditors survive in population 1. By $T=150$, in the region of the square lattice corresponding to the cluster of compliant firms, the best response for auditors now is not to inspect. But in that region all auditors' neighbours do inspect so they are not able to revert their strategy back to no inspection. Given the cluster of inspecting auditors moves faster than the cluster of compliant firms, at the boundaries between clusters in population 1 the best response of inspecting remains true. And the same happens with the best response to comply in population 2 until the evolutionary equilibrium is achieved $\bullet$

\textbf{Proposition 2:} For intermediate to high inspection cost, there is either a long run robust evolutionary equilibrium in which the pollution trap scenario disappears or an oscillatory long run behaviour after a transient period in which the shares of inspecting auditors and of compliant firms keep changing over time without reaching an equilibrium. Moreover, the likelihood of reaching an evolutionary equilibrium decreases with increasing cost of inspection. The long run average behaviour of each population does not match with that found in the well-mixed population evolutionary game, although the long run average values of $p$ and $q$ do follow the sensitivity analysis discussed in subsection \ref{sec:game}, i.e., $\frac{\partial\left(\pi_A^{I}-\pi_A^{N} \right) }{\partial c}<0\wedge\frac{\partial\pi_F^{P}}{\partial p}<0$: as the inspection cost increases, the likelihood of inspecting decreases, leading also to an increase in the share of non-compliant firms.

\begin{figure}[htp]
	\centering
	\begin{tabular}{cc}
		\epsfig{file=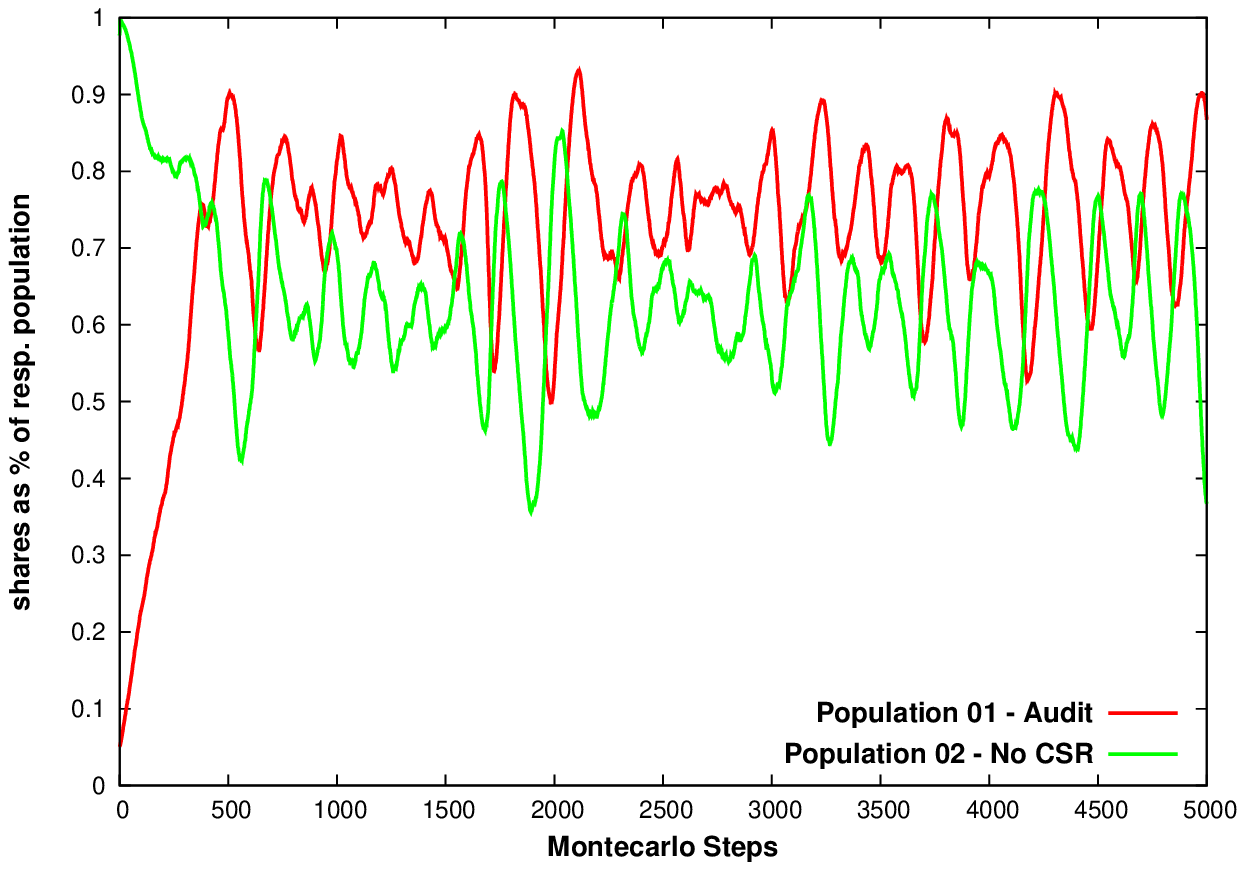,height=4.25cm,width=7.75cm,angle=0}&
		\epsfig{file=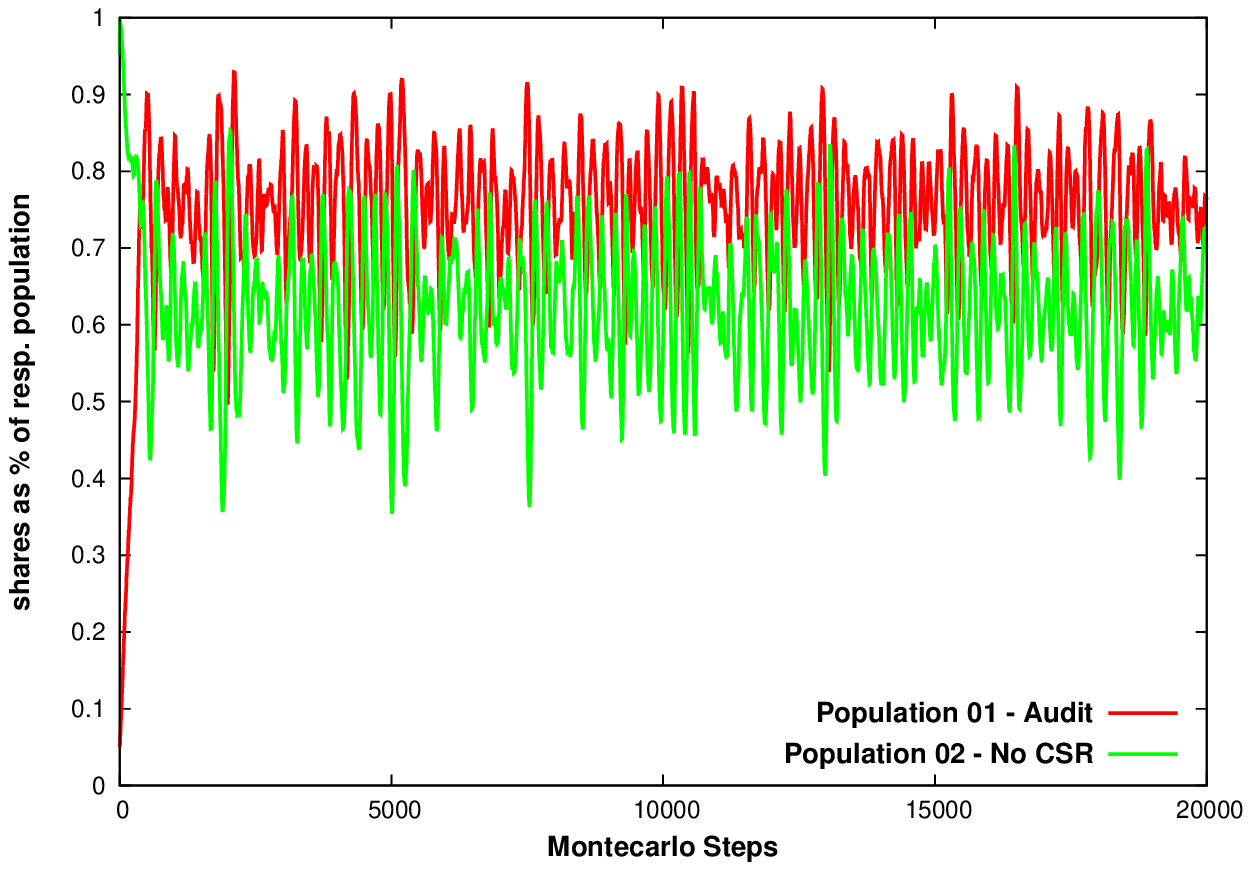,height=4.25cm,width=7.75cm,angle=0}
	\end{tabular}
	\caption{(color online) time evolution of the state of each population for $k=1$, $\bar{\beta}=2$, $\xi=0.75$ and $(p_0=0.05,q_0=0.95)$.}
	\label{fig:5}
\end{figure}

\textbf{Discussion:} The numerical simulations displayed possible oscillatory behaviour for $0.60\leq\xi\leq 0.90$. Out of 50 realizations performed with different seeds for each value of $\xi$, oscillatory long run behaviour was registered 4 ($\xi=0.60$), 3 ($\xi=0.65$), 31 ($\xi=0.75$) and 50 ($\xi=0.90$) times. Thus, as the cost of inspection increases, it reaches a boundary above which a long run equilibrium is not possible. In Figure \ref{fig:5} the evolution of both populations over time is displayed for the case $\xi=0.75$ showing the first 20,000 MCS and just the first 5,000 MCS as well. From the latter, it can be seen that after 500 MCS the non-oscillatory transient period is already over. Similar evolutionary patterns were found for $\xi=0.60$, $\xi=0.65$ and $\xi=0.90$. 

\begin{figure}[htp]
	\centering
	\begin{tabular}{cc}
		\epsfig{file=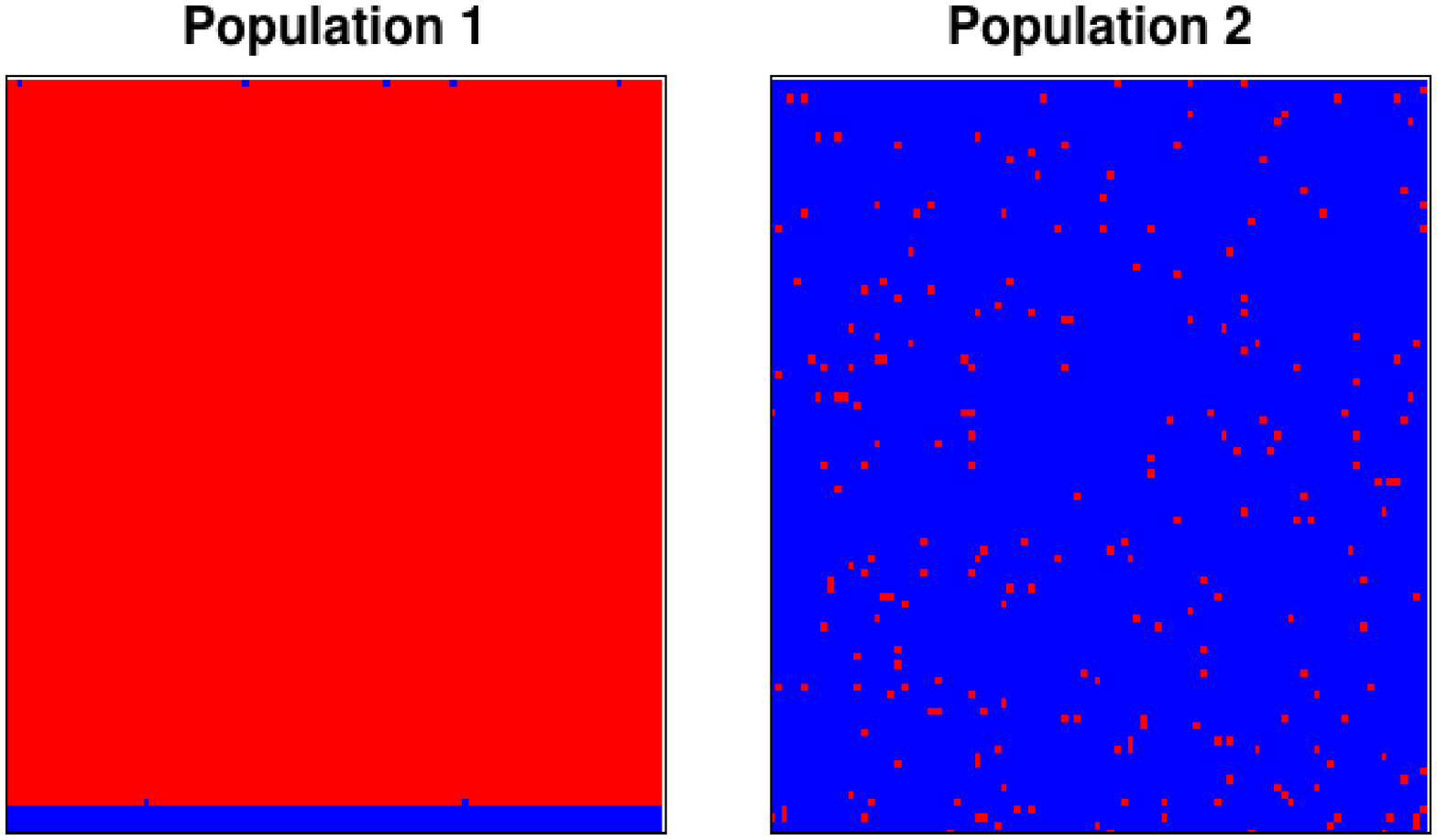,height=3.6cm,width=8cm,angle=0}&
		\epsfig{file=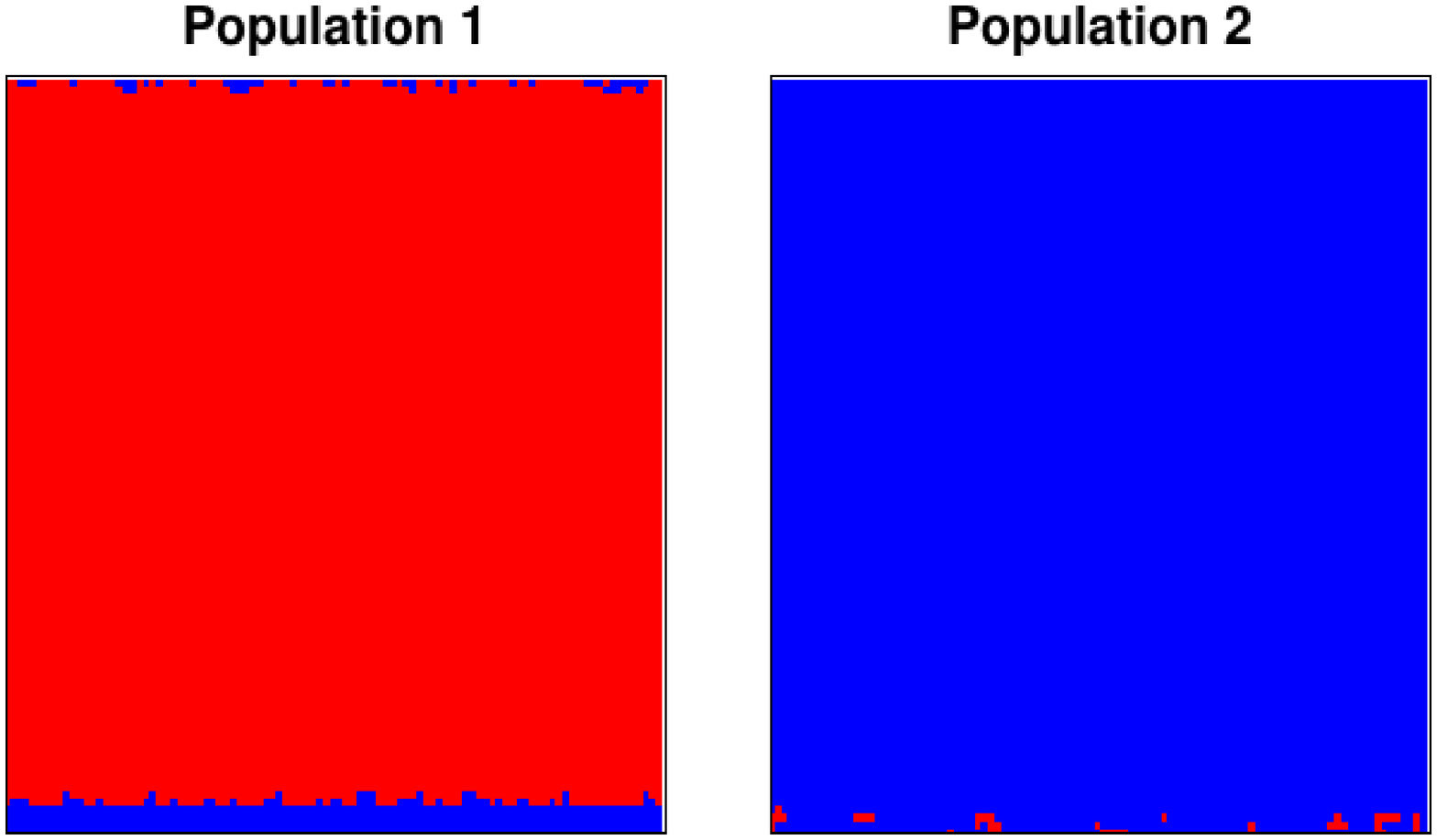,height=3.6cm,width=8cm,angle=0}\\
		\epsfig{file=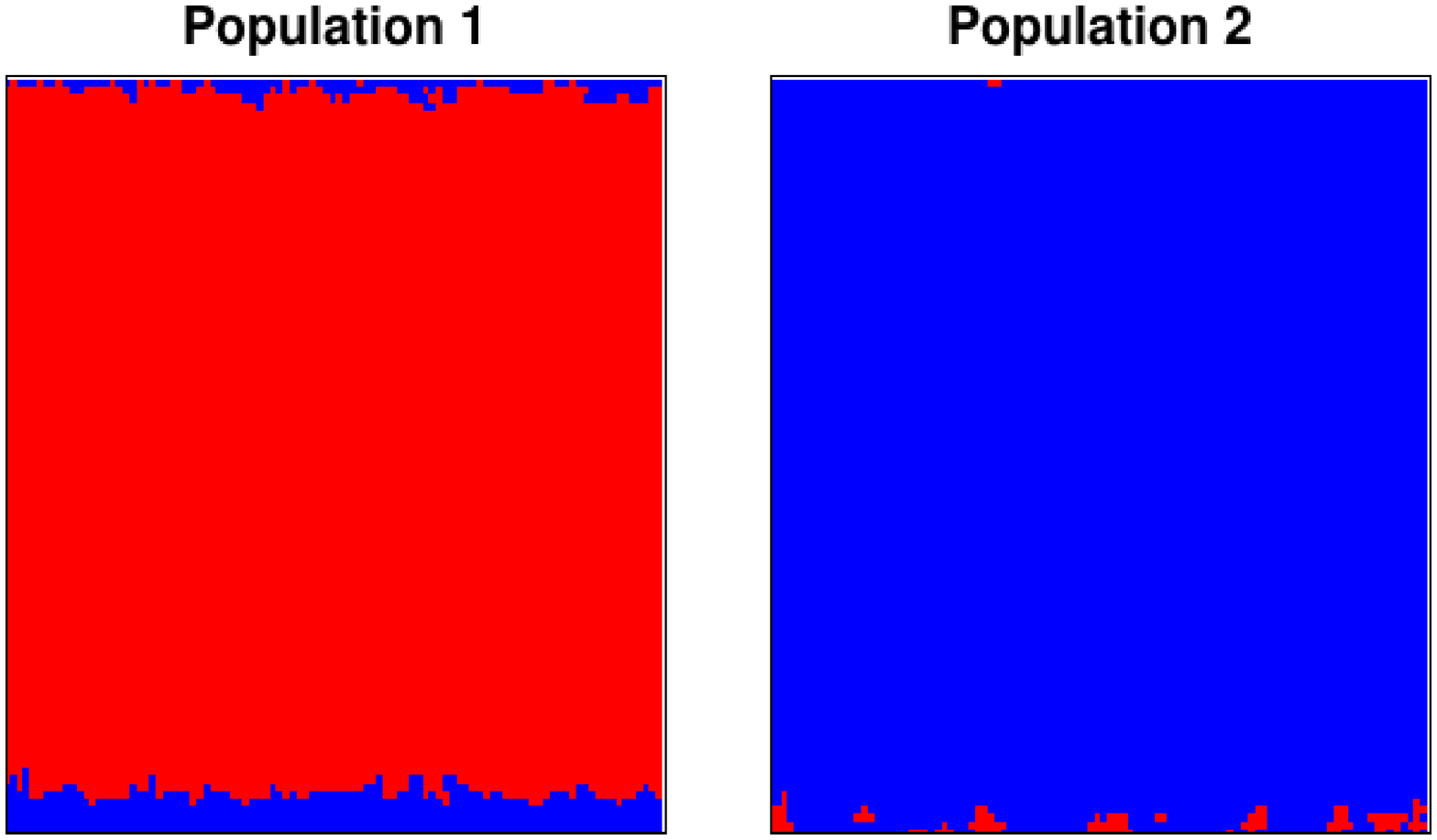,height=3.6cm,width=8cm,angle=0}&
		\epsfig{file=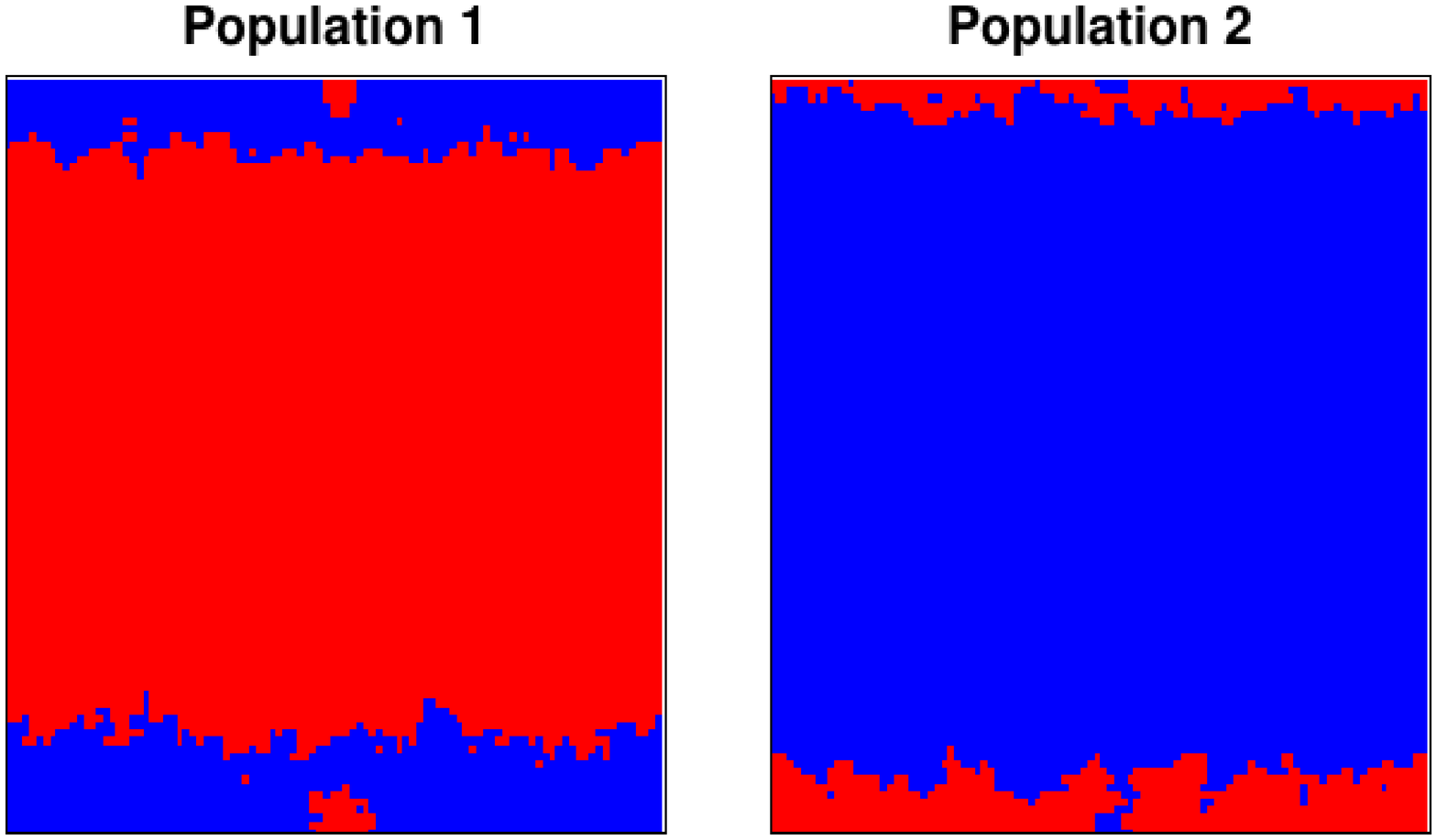,height=3.6cm,width=8cm,angle=0}\\
		\epsfig{file=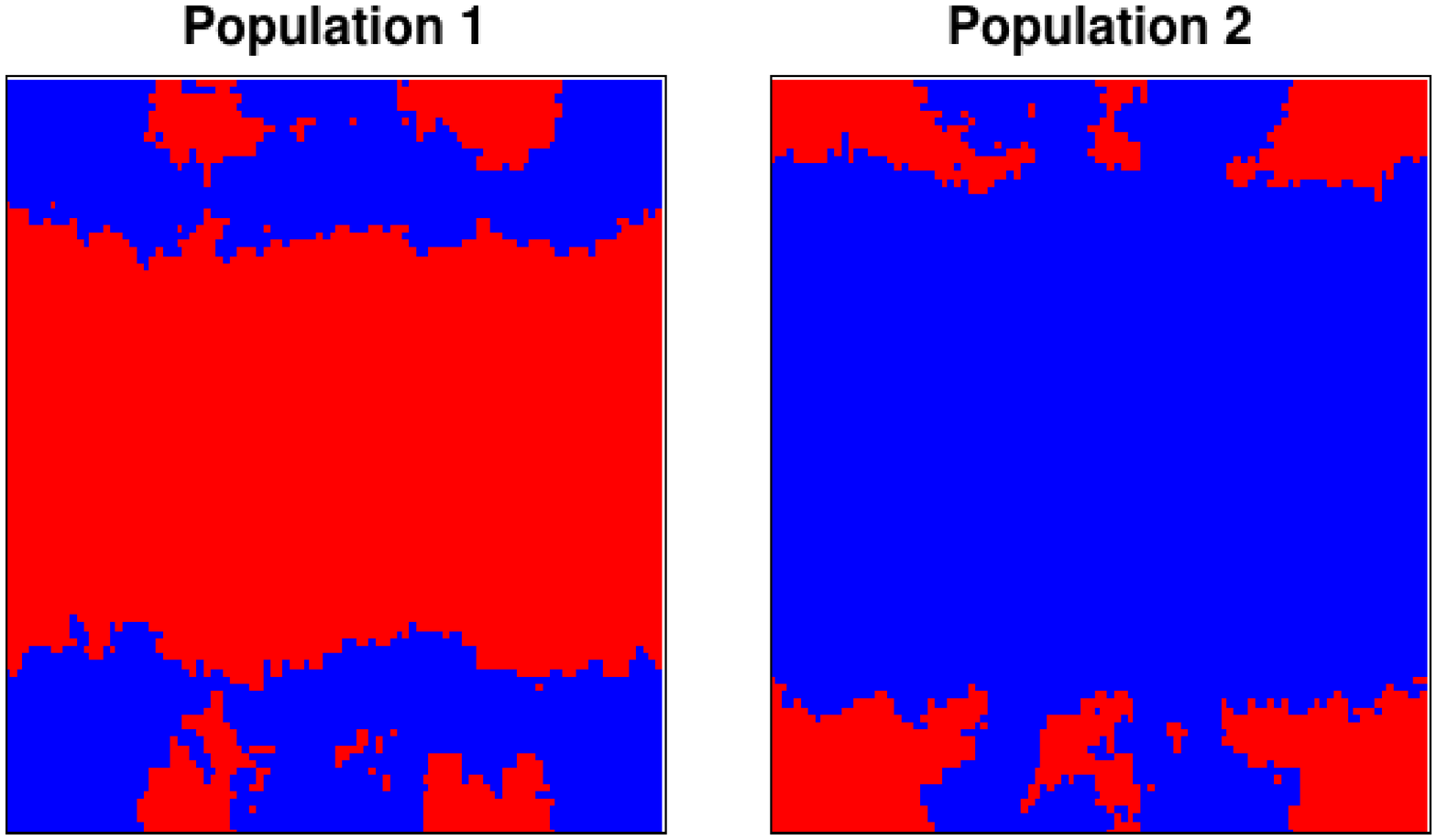,height=3.6cm,width=8cm,angle=0}&
		\epsfig{file=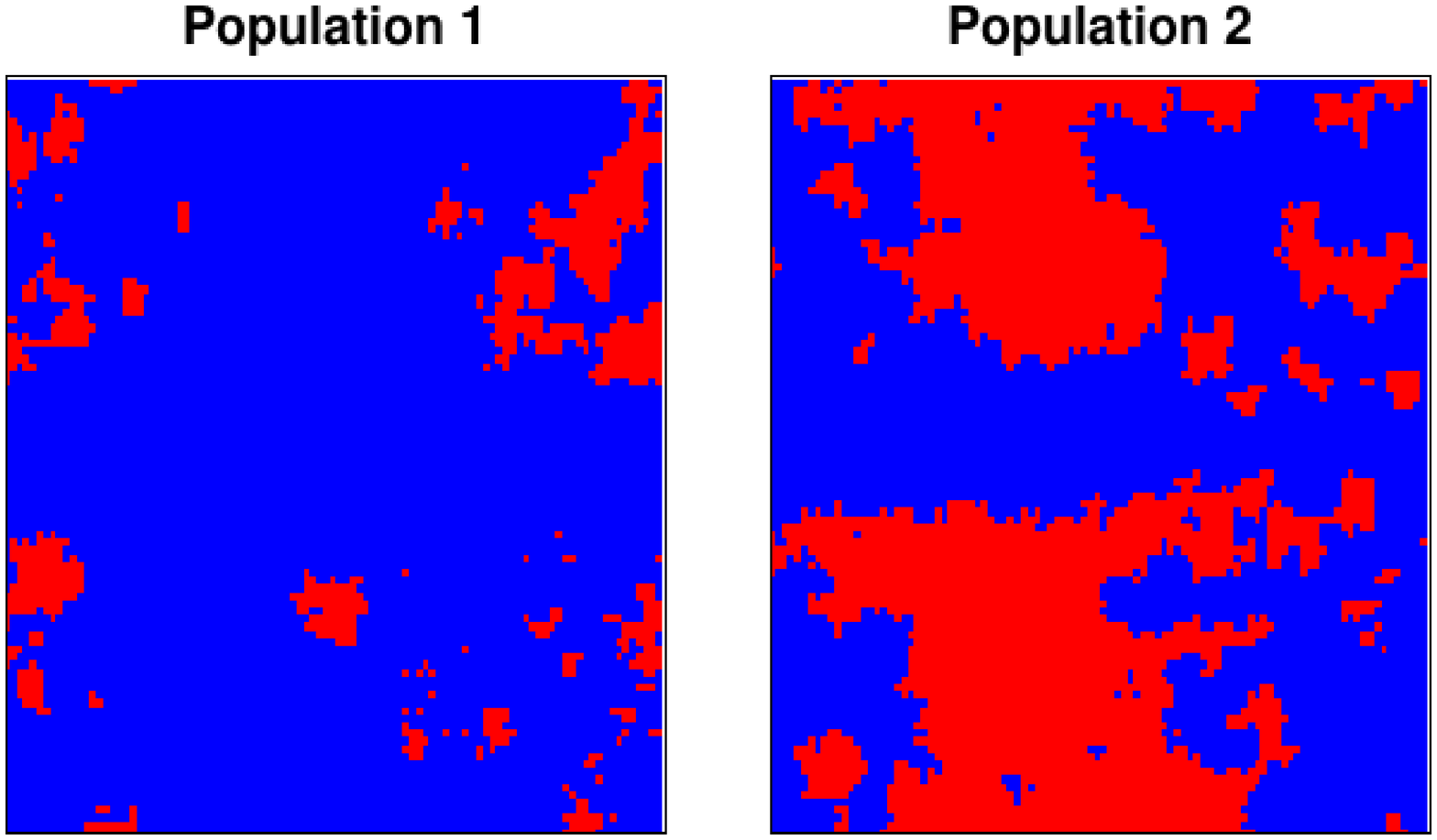,height=3.6cm,width=8cm,angle=0}\\
		\epsfig{file=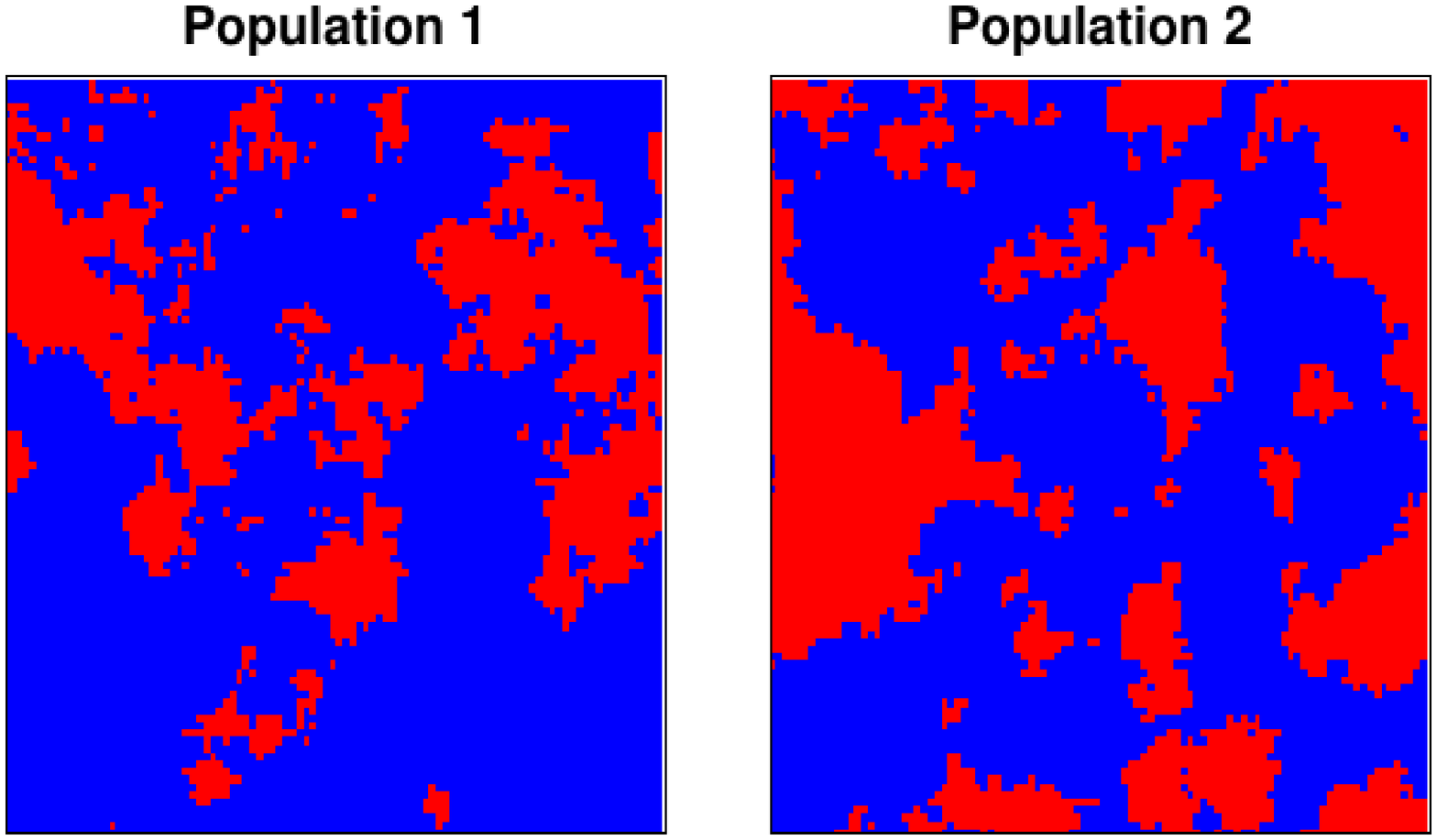,height=3.6cm,width=8cm,angle=0}&
		\epsfig{file=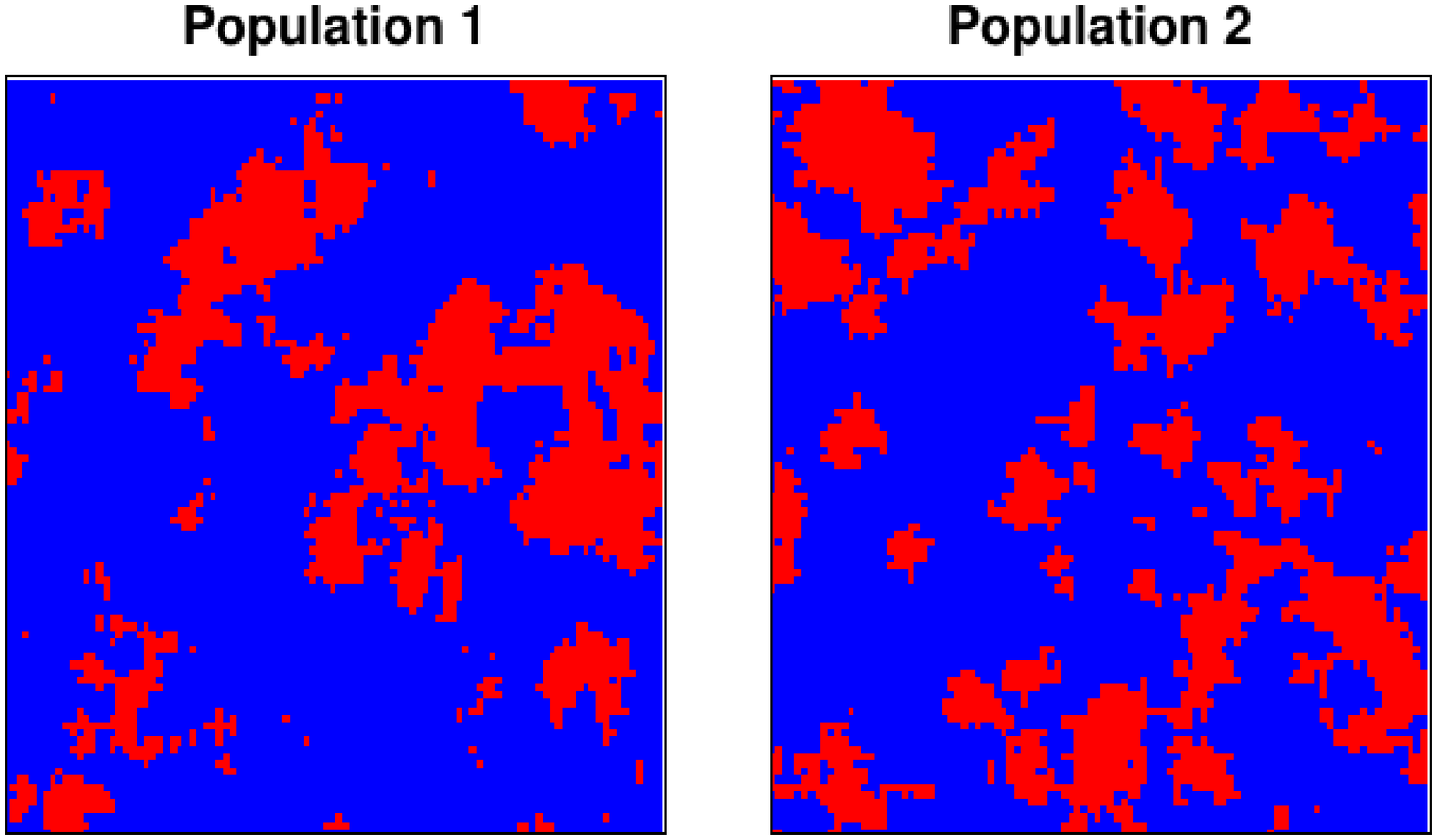,height=3.6cm,width=8cm,angle=0}
	\end{tabular}
	\caption{(color online) evolution of the state of each population for parameters: $k=1$, $\bar{\beta}=2$, $\xi=0.75$ and initial conditions $(p_0=0.05,q_0=0.95)$. In both populations, blue (resp. red) corresponds to strategy 1 (resp. 2). From top left to bottom right, square lattice at $T=1;8;22;100;200;500;8,000;20,000$.}
	\label{fig:6}
\end{figure}

In Figure \ref{fig:6} we present the evolution of the $L\times L$ square lattice for a particular realization when oscillation occurs, using $k=1$, $\bar{\beta}=2$, $\xi=0.75$ and initial conditions $p_0=0.05\wedge q_0=0.95$ (pollution trap), from Montecarlo step $T=1$ to $T=20,000$. When compared to a realization without oscillation, the main important difference is that, when the cluster of inspecting auditors starts to spread in population 1, the spreading speed is not as fast as the speed in realizations with lower costs of inspection (e.g.: compare the size of the blue strip in population 1 for $T=22$ in Figure \ref{fig:6} with that for $T=17$ in Figure \ref{fig:2}). This happens because, although the best response for auditors is still to inspect, the difference in profits now is smaller. Then, when inspecting auditors compare their payoffs with non-inspecting auditors in their neighbourhood, although the former tend to perform better, the strategy switching process is probabilistically increasing in the payoff difference. Hence, the switching process now becomes slower than when the inspection cost was lower. With the clusters of compliant firms starting to grow in population 2, non-inspecting auditors at the boundary between clusters are now able to re-invade that region and create a small compact cluster (see $T=100$). Once this happens, convergence to an evolutionary equilibrium with one strategy being driven to extinction is each population is doomed. Clusters of different strategies keep moving around the square lattice in both populations due to invasion and re-invasion with the state of both populations oscillating over time.

\begin{figure}[H]
	\centering
	\begin{tabular}{cccc}
		\epsfig{file=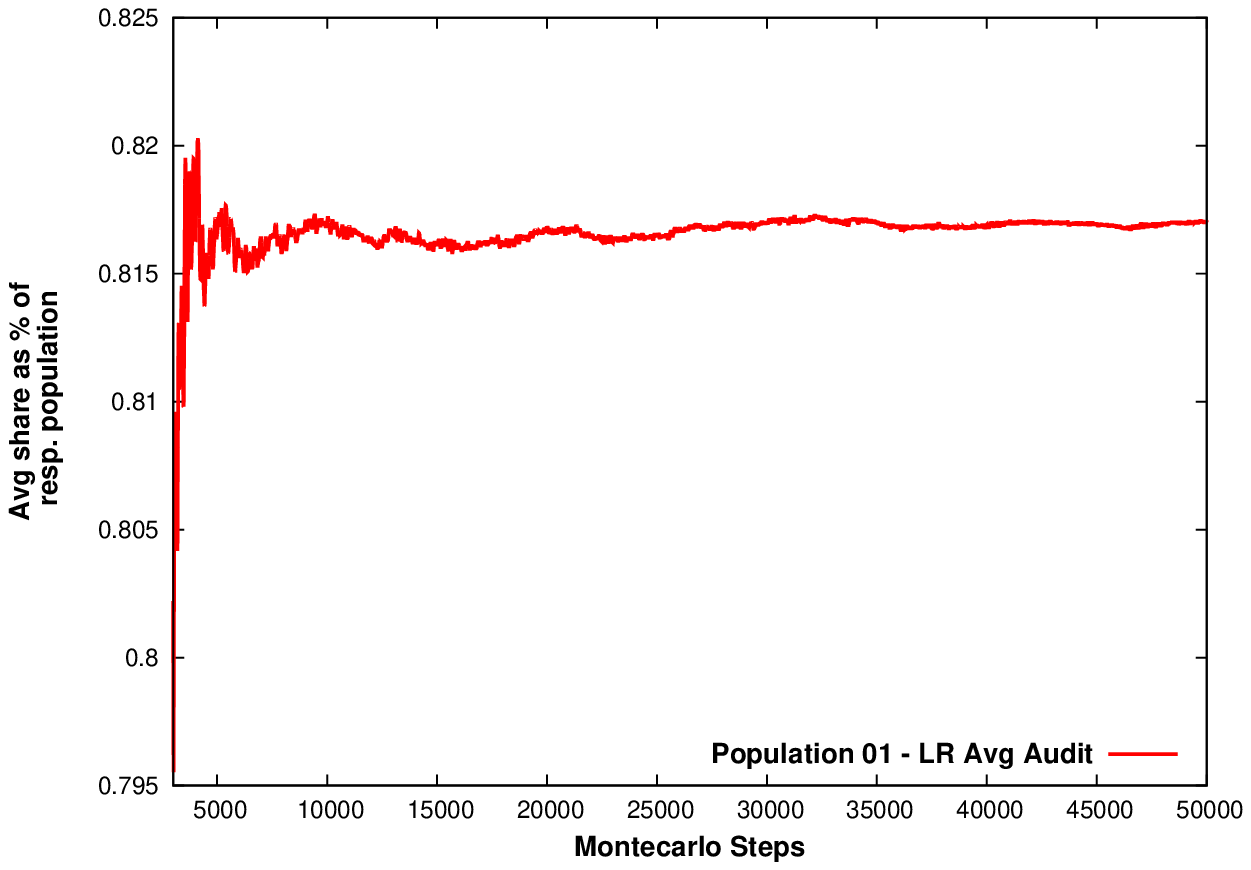,height=3.7cm,width=7.75cm,angle=0}&
		\epsfig{file=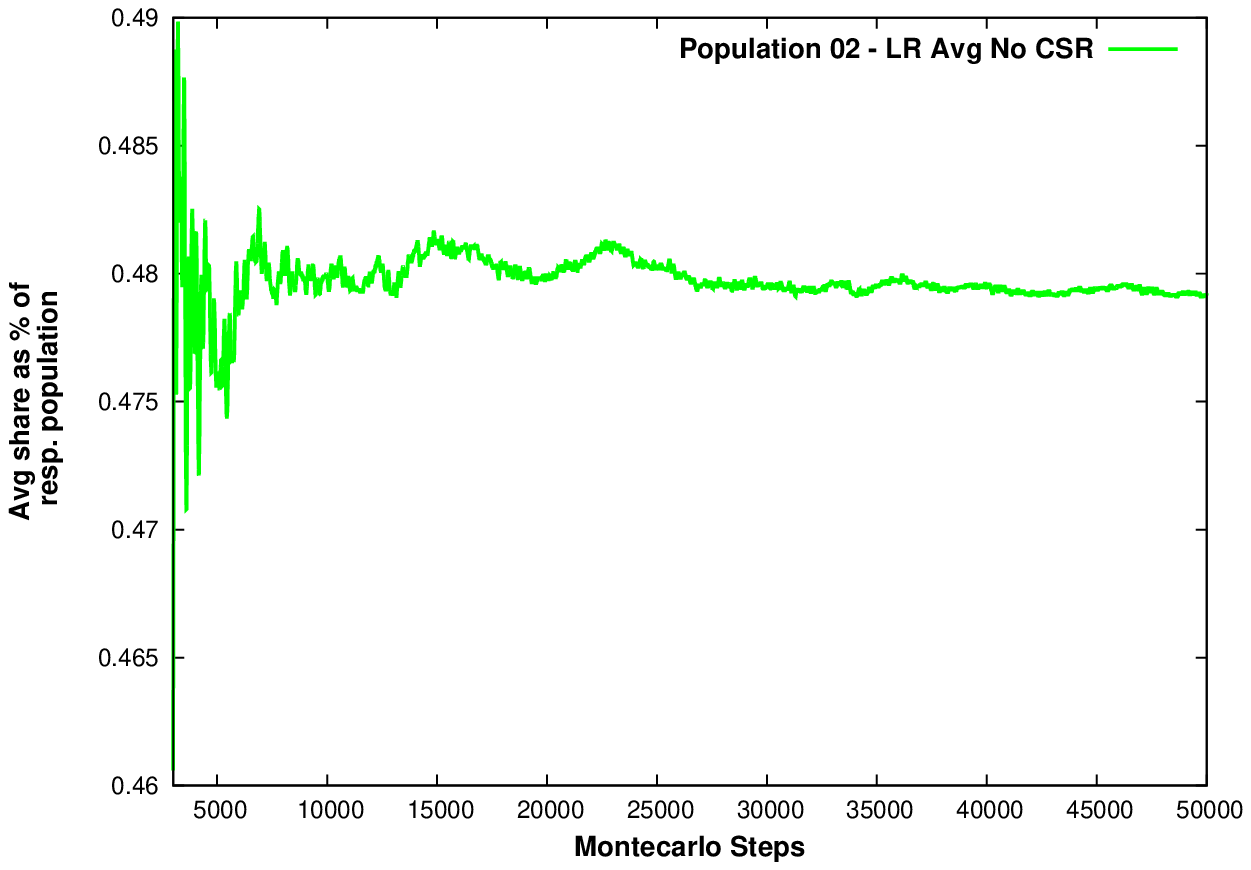,height=3.7cm,width=7.75cm,angle=0}\\
		\epsfig{file=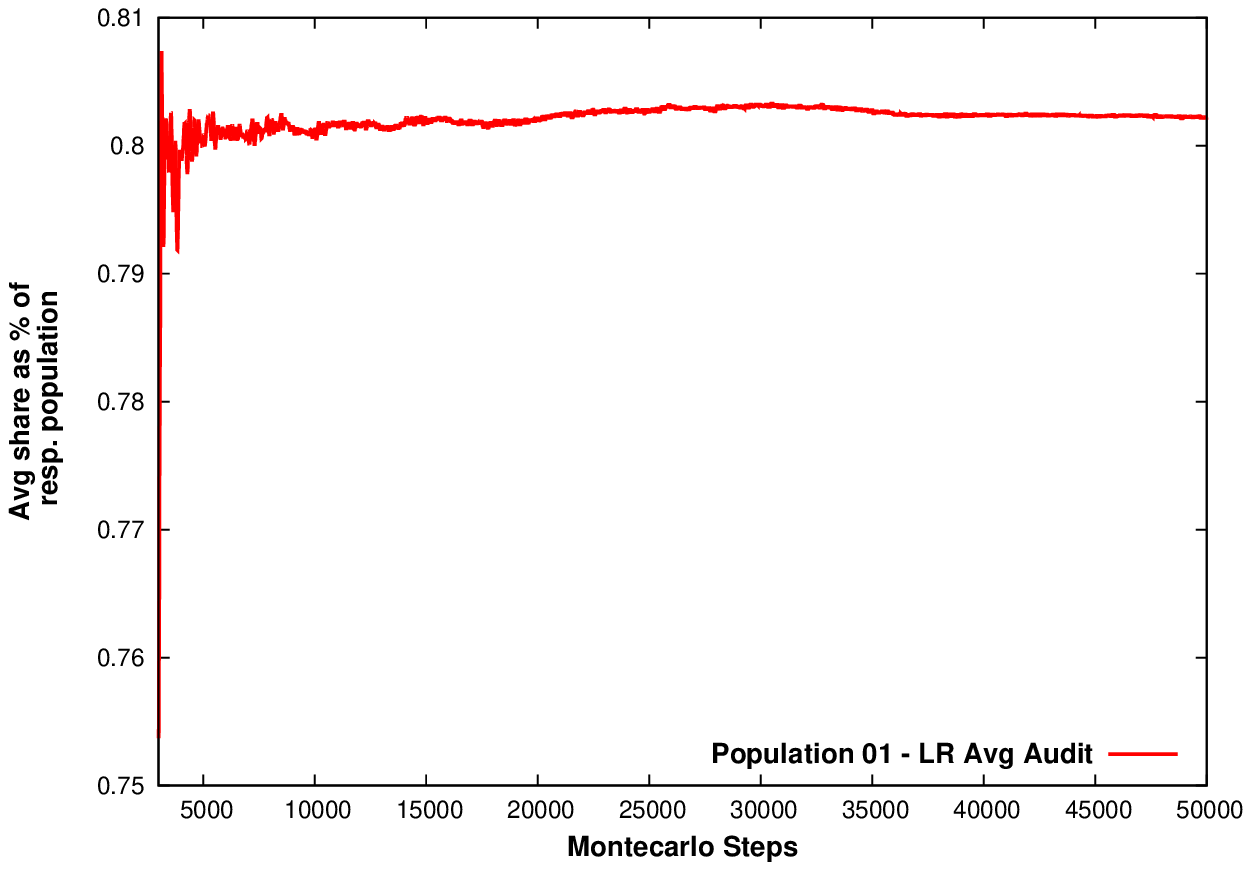,height=3.7cm,width=7.75cm,angle=0}&
		\epsfig{file=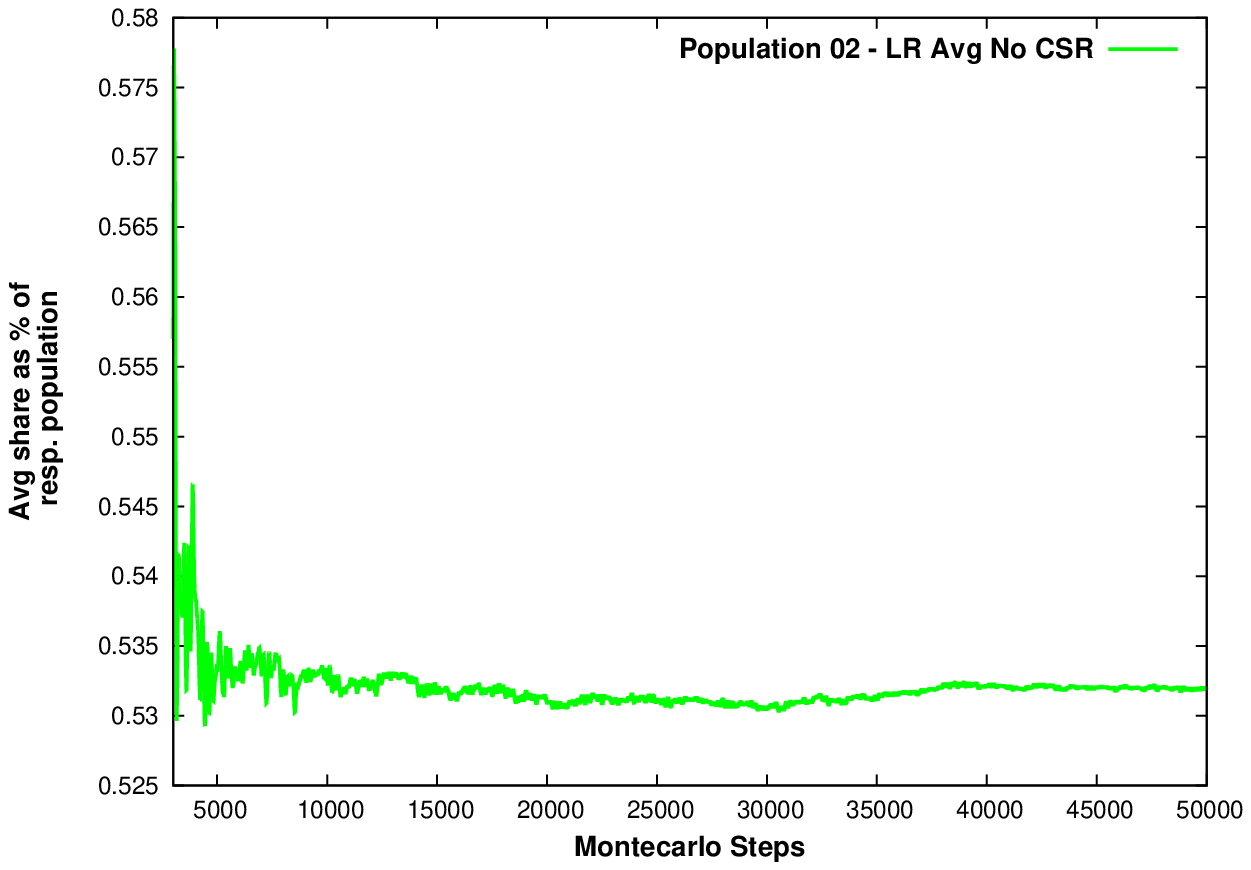,height=3.7cm,width=7.75cm,angle=0}\\
		\epsfig{file=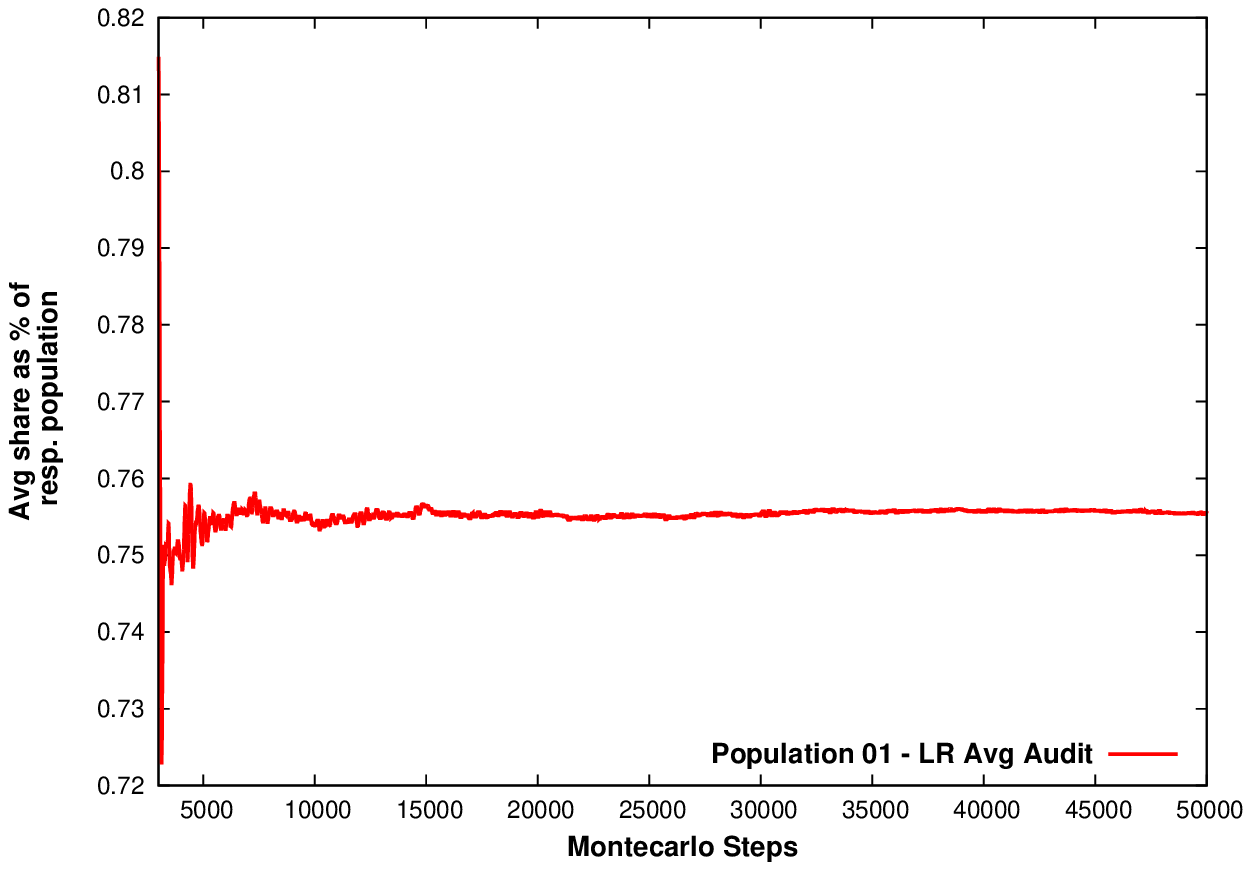,height=3.7cm,width=7.75cm,angle=0}&	
		\epsfig{file=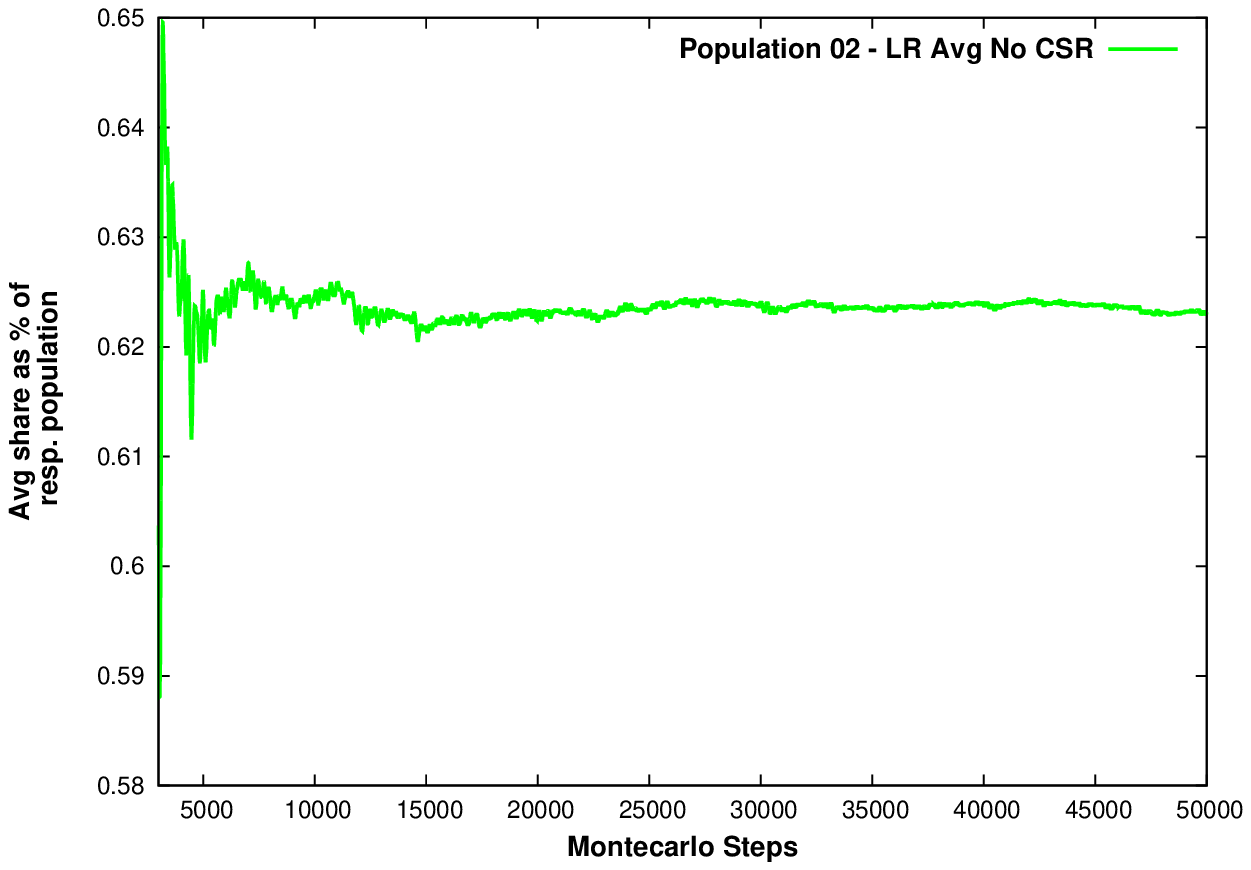,height=3.7cm,width=7.75cm,angle=0}\\
		\epsfig{file=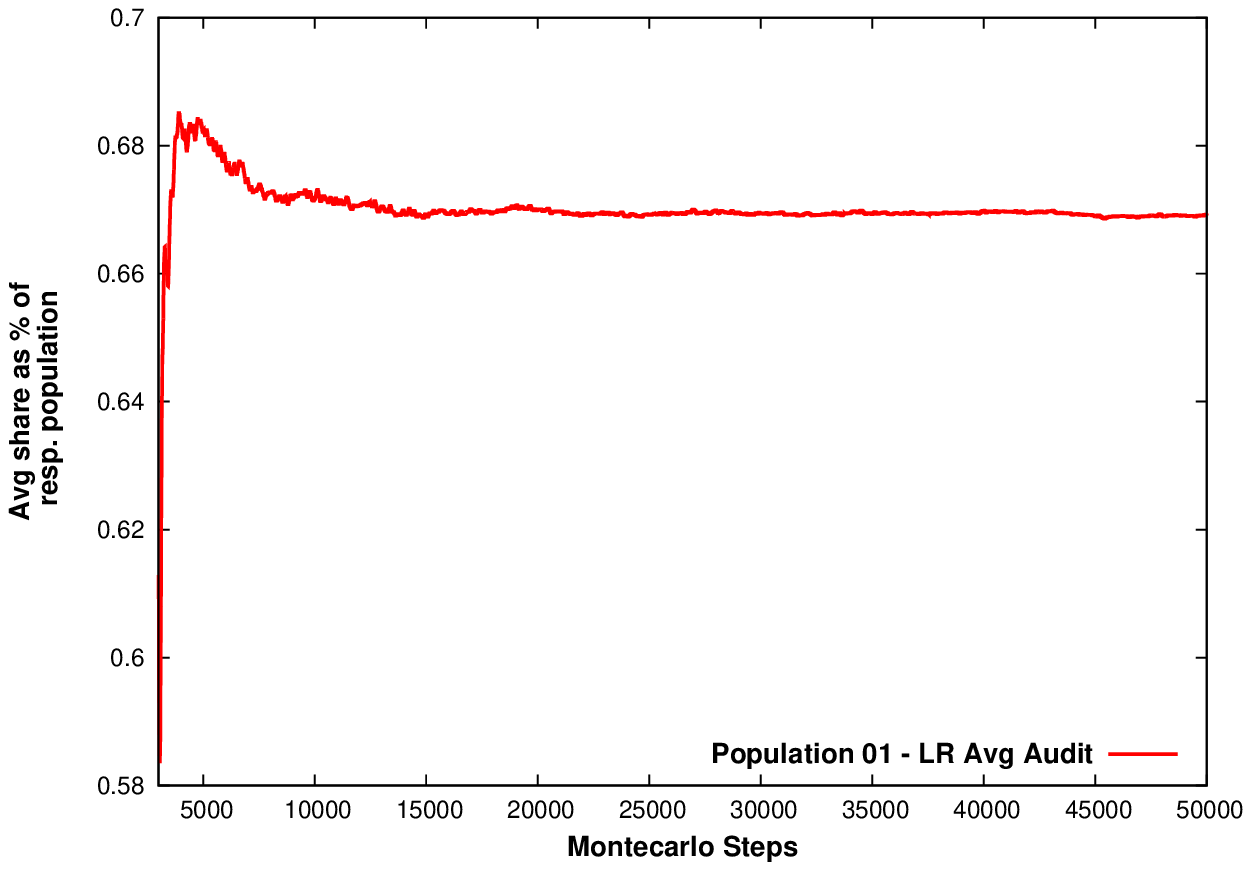,height=3.7cm,width=7.75cm,angle=0}&
		\epsfig{file=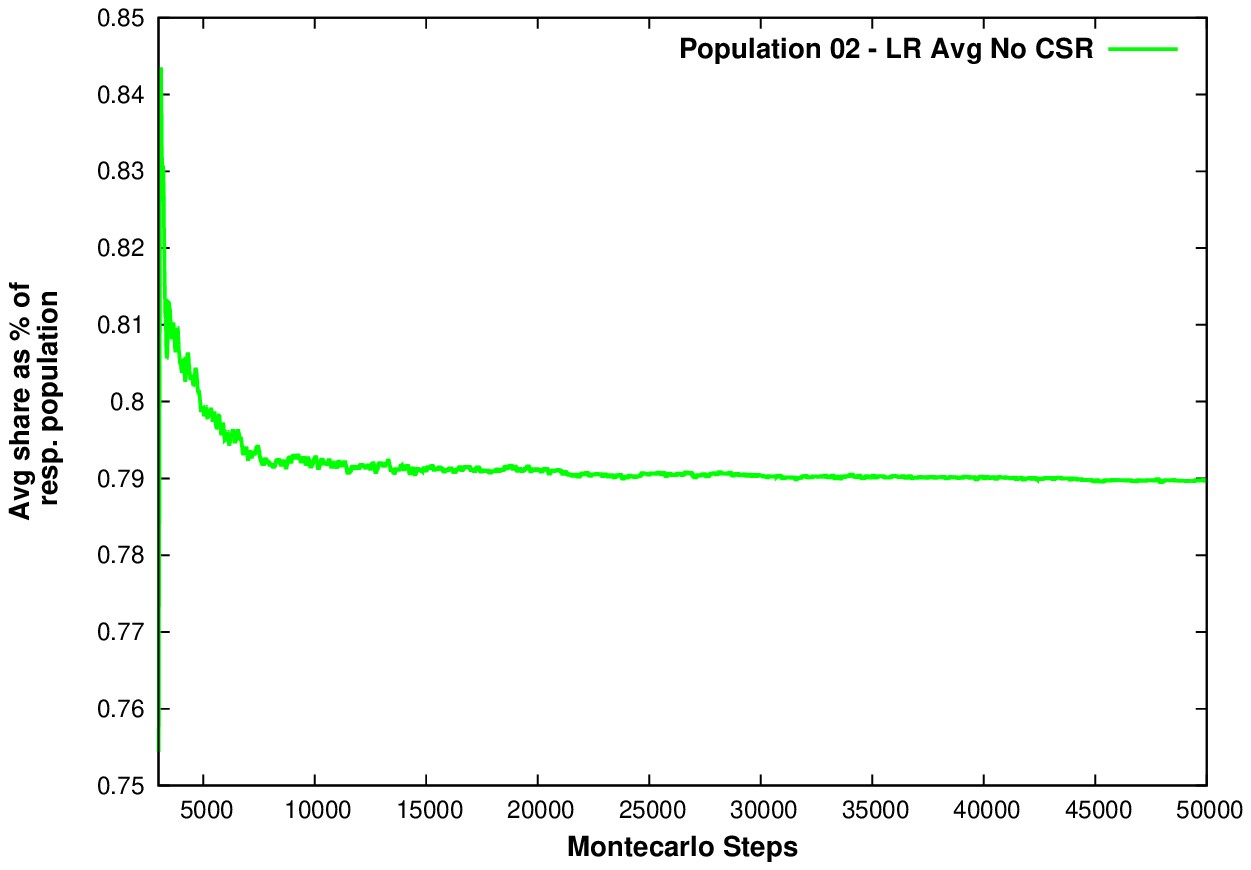,height=3.7cm,width=7.75cm,angle=0}
	\end{tabular}
	\caption{(color online) long run average state of each population for $k=1$, $\bar{\beta}=2$ and $(p_0=0.05,q_0=0.95)$ for MCS from $T=3,000$ to $T=50,000$. From top to bottom, $\xi=0.60$; $\xi=0.65$; $\xi=0.75$; $\xi=0.90$.}
	\label{fig:4}
\end{figure}

\begin{figure}[H]
	\centering
	\begin{tabular}{cccc}
		\epsfig{file=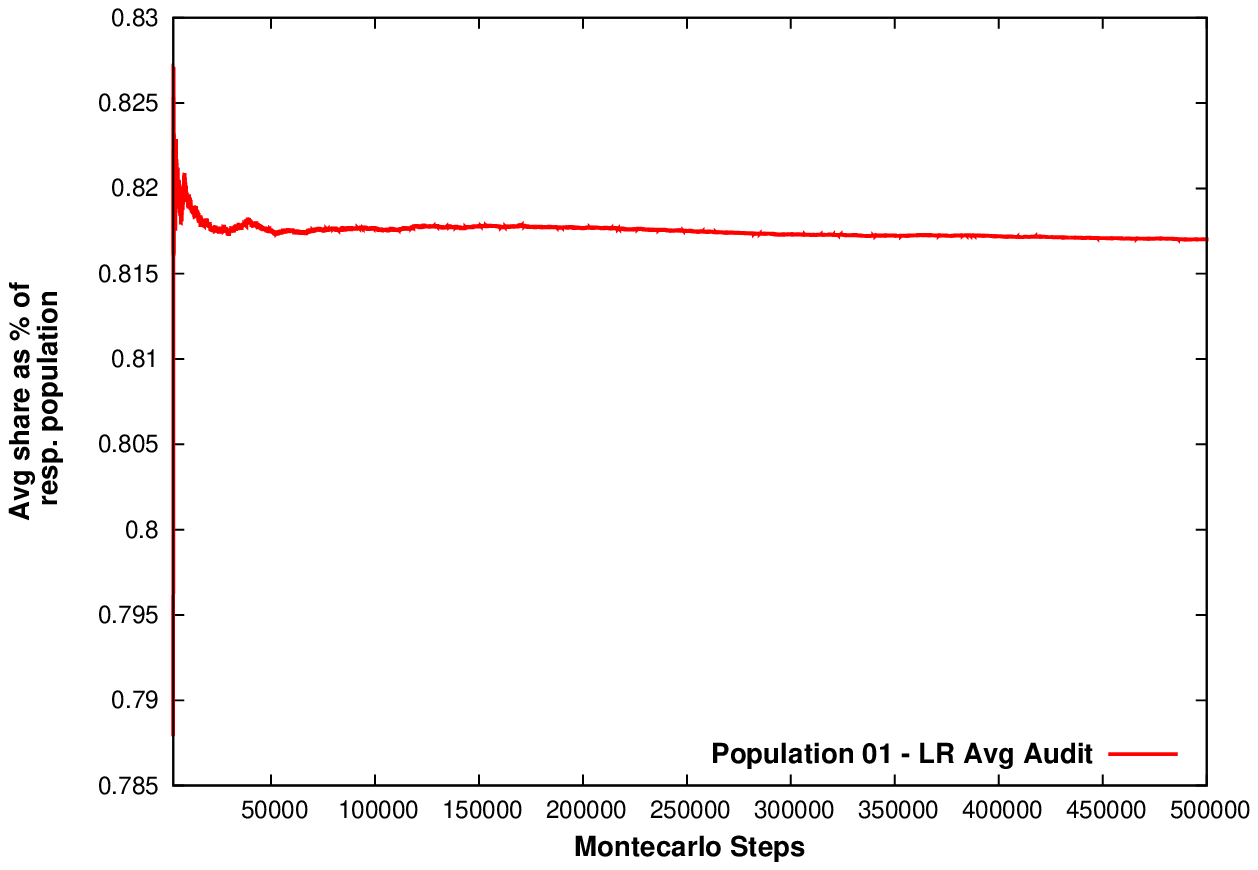,height=3.5cm,width=7.75cm,angle=0}&
		\epsfig{file=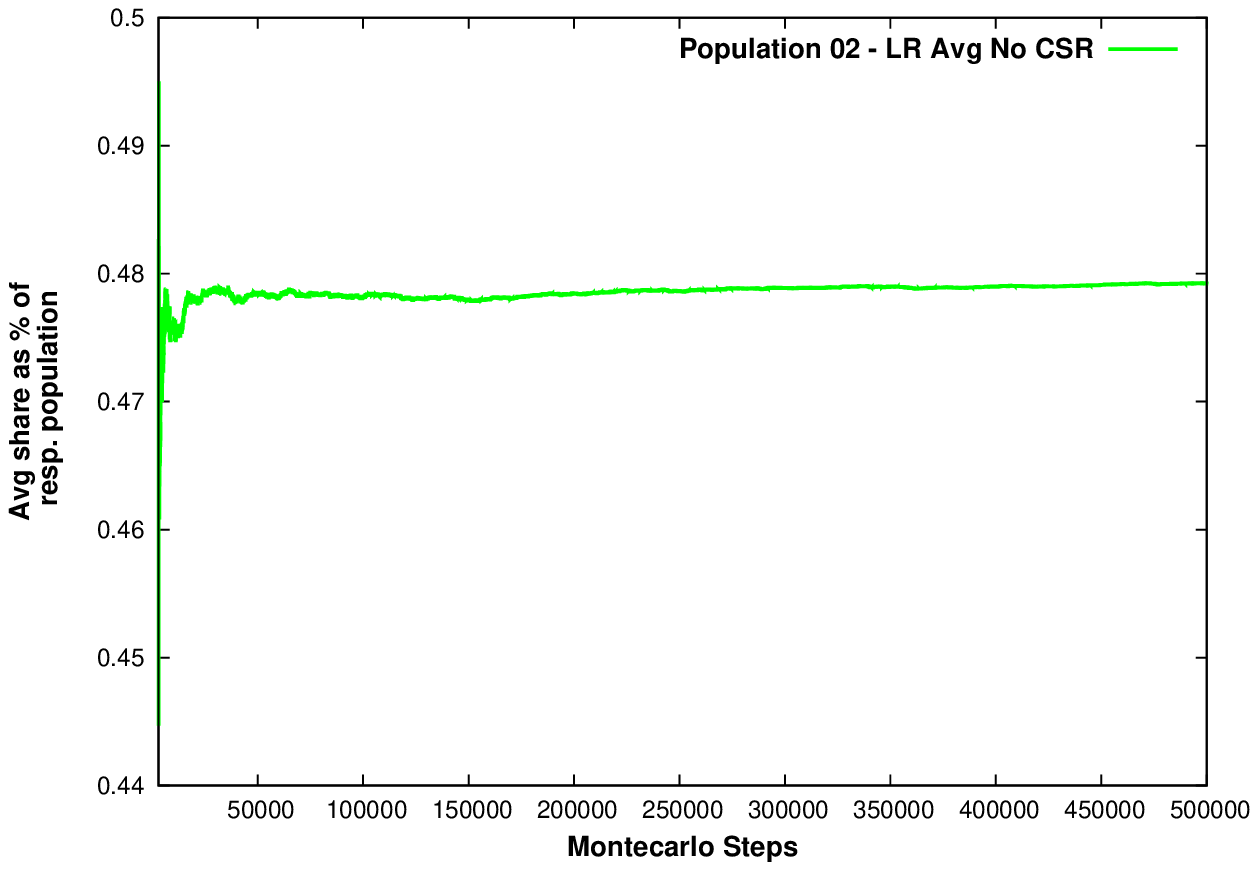,height=3.5cm,width=7.75cm,angle=0}\\
	\end{tabular}
	\caption{(color online) long run average state of each population for $k=1$, $\bar{\beta}=2$, $\xi=0.60$ and $(p_0=0.05,q_0=0.95)$ for MCS from $T=3,000$ to $T=500,000$.}
	\label{fig:8}
\end{figure}

In order to understand the long run average behaviour of each population, simulations with 50,000 MCS were carried out for each value of $\xi$ and we computed the average state of each population after 3,000 MCS, updating the computation step after step until 50,000 MCS, in order to guarantee the transient period was over and had no effect on the computed average states. In Figure \ref{fig:4} the long run average shares of inspecting auditors in population 1 and of non-compliant firms in population 2 are displayed. Clearly, as the cost of inspection increases, the long run proportion of inspecting auditors decreases and the proportion of non-compliant firms increases. In order to ensure that 50,000 MCS were more than enough to capture the long run average behaviour of each population, we repeated the same procedure but with 500,000 MCS for the case with $\xi=0.60$. Results are displayed in Figure \ref{fig:8} and one can see a very good level of agreement between averaged values of both $p$ and $q$ when compared to those in Figure \ref{fig:4} using only 50,000 MCS $\bullet$\\

\textbf{Proposition 3:} When the inspection cost approaches $\bar{c}$, the oscillatory evolutionary pattern ceases to occur and is replaced by a different long run evolutionary equilibrium in which $p^*=1$ and $q^*=1$, i.e., a country where all auditors carry out a very costly and ineffective inspection process that does not lead to corporate environmental compliance. Moreover, the few compliant firms that were located in the country are driven to extinction.

\begin{figure}[H]
	\centering
	\begin{tabular}{cc}
		\epsfig{file=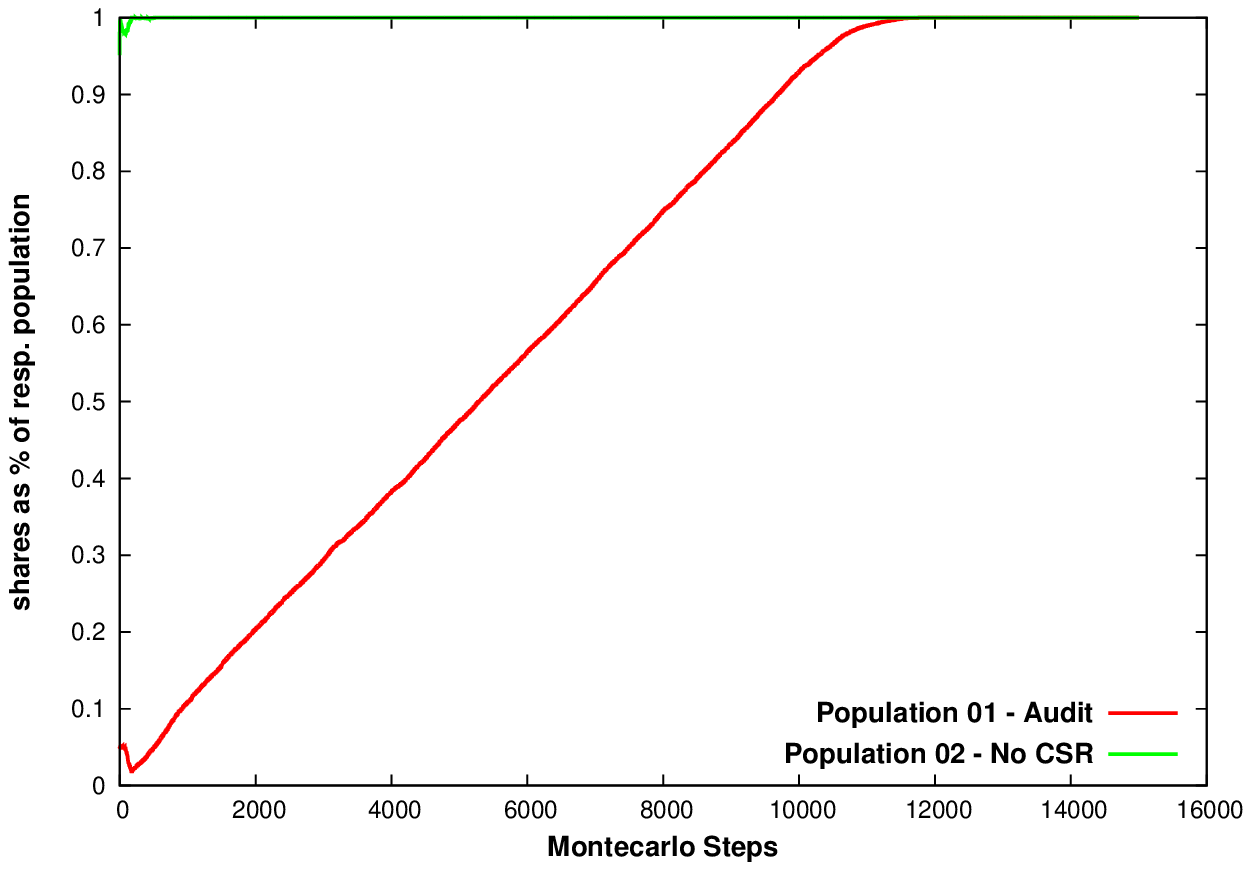,height=3.5cm,width=7.75cm,angle=0}&
		\epsfig{file=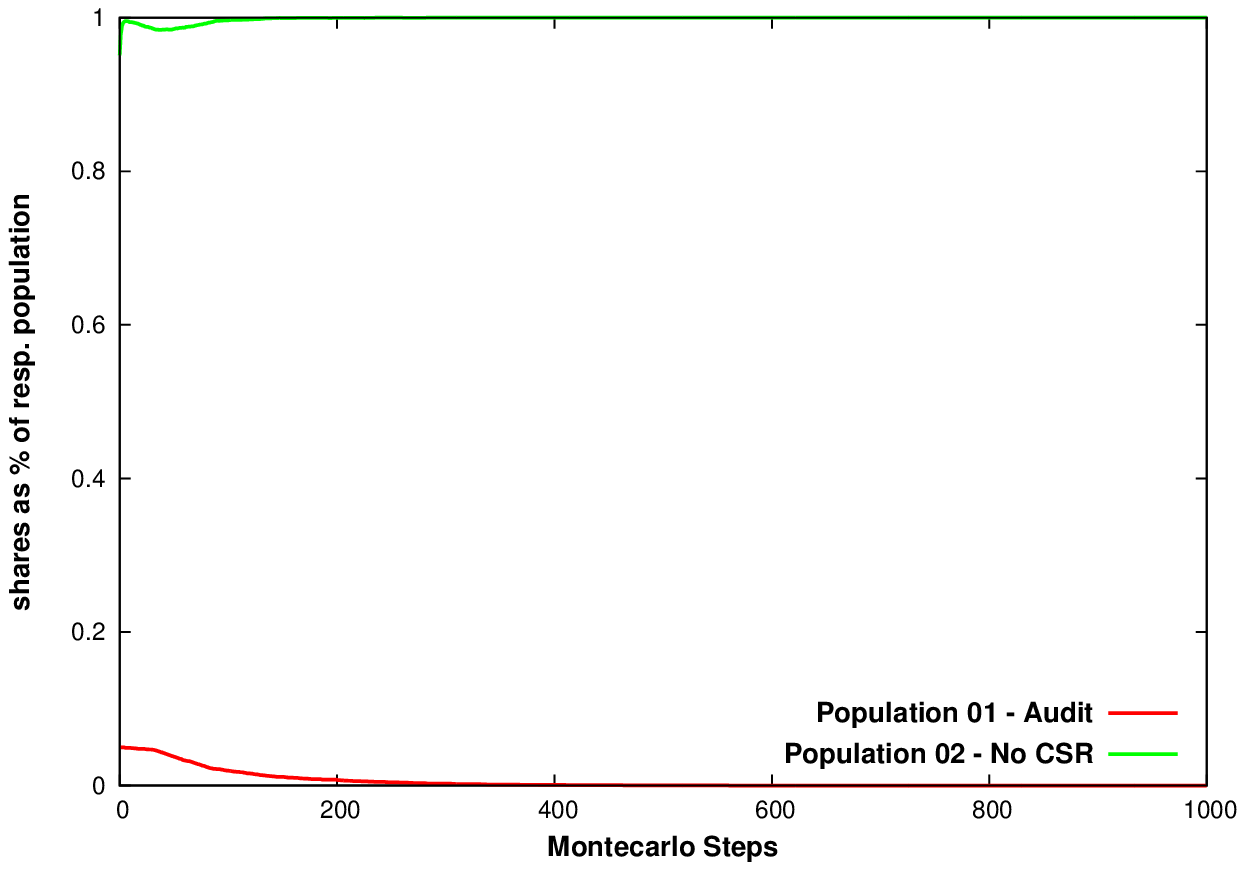,height=3.5cm,width=7.75cm,angle=0}
	\end{tabular}
	\caption{(color online) time evolution of the state of each population for $k=1$, $\bar{\beta}=2$, $(p_0=0.05,q_0=0.95)$, $\xi=0.99$ (left panel) and $\xi=1.02$ (right panel).}
	\label{fig:7}
\end{figure}

\textbf{Discussion:} As in Propositions 1 and 2, the inspecting cluster in the population of auditors starts to grow but now the growth speed is so slow that in population 2 the cluster of compliant firms in the same region grows at a similar speed. Consequently, non-inspecting auditors are able to re-invade the small strip to an extent that the growth of compliant firms slows down and becomes negative until all compliant firms disappear from the population. Once this happens, the cluster with inspecting auditors starts to increase again at a very slow speed until taking over population 1, but without any impact on compliance given all compliant firms were driven out of the country. The evolution of the state of both populations is displayed in the left panel of Figure \ref{fig:7} $\bullet$\\

Increasing $c$ further, $c>\bar{c}$ and the stage game in (\ref{Matrix-II}) has one unique Nash equilibrium in pure strategies $p^*=0$ and $q^*=1$, i.e., a pollution trap. Within this range of values in which $\xi>1$, the numerical simulations showed that the only evolutionary equilibrium in both the well-mixed population and the spatial game do match the classic game Nash equilibrium. The right panel in Figure \ref{fig:7} shows the evolution of both populations for the case $\xi=1.02$.

Based on the long run behaviour of population 2 displayed in Figures \ref{fig:4}-\ref{fig:7}, except for the case when $\xi\rightarrow 1$, the spatial grid favoured environmental compliance given that results showed a long run average value of $q$ below the one found in the mean-field approximation, i.e., for $\xi\in\left[0.05;0.90 \right]$, $\bar{q}_{\text{SPATIAL}}<\bar{q}_{\text{MEAN-FIELD}}\equiv\xi$.
 
\section{Conclusion}
\label{sec:conclusion}
We studied the interplay between corporate environmental compliance and policy regulation in a context of a competitive market where enforcement requires costly inspections and there is a trade-off between maximizing social welfare and maximizing private profits. Firms are better off whenever they are non-compliant with environmental regulations and they do not suffer any form of inspection. On the other hand, social welfare is maximized whenever the policy maker does not need to enforce inspection and firms have a corporate culture of internalizing the negative externality due to pollution when making their production decisions. Such decisions take into account both the type of production technology they use as well as the level of emission. We assumed a society in a pollution trap scenario, i.e., a scenario in which the vast majority of firms are not environmental friendly and auditors have a culture of not inspecting firms.

An evolutionary game was used to model such trade-off involving the conflict of interests between firms and the policy maker. All players were bounded rational such that, despite being able to compute, maximize and compare profits given their decision variables, they were not sure which strategy was the best in the long run. We started with a benchmark model assuming that populations were well-mixed, i.e., firms were neighbours to each other and could also be inspected by any randomly designated auditor. This approach was in line with previous evolutionary game models found in the literature such as in Anastasopoulos and Anastasopoulos (2012), Cressman et al. (1998) and Xiao and Yu (2006), among others. Using replicator dynamics to model the game, results did not find a long run evolutionary equilibrium. Instead, there was an oscillatory evolutionary pattern in which the state of each population kept changing over time, similar to the pattern found in the Lotka-Volterra model.

We then extended the benchmark model by assuming the two populations were allocated in a square grid, with each location being shared by a firm and an auditor. This is more in line with the real world where firms are geographically spread and auditors tend to work close to where they live. The extended model displayed a quite rich dynamics in which not only cyclical evolution was still possible but robust evolutionary equilibria could exist, depending on the cost of inspection. When the latter was very close to the maximum cost of inspection, the extended model showed that there is no advantage in the auditing process. The latter not only is not able to remove society from the pollution trap scenario, but it also leads the very few compliant firms located in the country to extinction.

Further research could take into account the impact of different degrees of environmental compliance across countries as well as the impact of internationalization on the corporate culture of firms. This could be modelled using a multilayer network as described in da Silva Rocha (2017). An initial idea could consider a two-layer network in which each layer would be an $L\times L$ square lattice with periodic boundary conditions. Each layer would correspond to a different country and each corresponding site of both layers would have an international subsidiary of each firm. Layers would be physically disconnected and interdependence would be only via the payoffs earned by firms from playing the evolutionary game against auditors in each layer. Auditors would be part of two independent populations, one in each layer, whose states would differ depending on the degree of inspection commitment. As in our model, a country in a pollution trap scenario would correspond to a layer with a residual proportion of auditors playing the strategy Inspect. Strategies used by players in each layer would be updated independently by comparing each player's payoff with that of a randomly chosen neighbour located in the corresponding layer. The main difference when compared to our model would be the fact that in each layer firms would compute their payoffs as a weighted average of the payoffs earned by each subsidiary (see Wang et al. 2012). A firm $X$ would compute the overall private profit of its subsidiary located in layer $x$ as $\Pi_x =\alpha\pi_x +(1-\alpha)\pi_x^{'}$, where $\pi_x$ (resp. $\pi_x^{'}$) would be the local profit earned by its subsidiary playing the game in layer $x$ (resp. $x^{'}$). Similarly, for its subsidiary located in layer $x^{'}$, the overall private profit would be $\Pi_x^{'} =(1-\alpha)\pi_x^{'}+\alpha\pi_x$. Parameter $\alpha\in[0,1]$ would measure the bias in the consideration of payoffs collected in the two layers, where the bias could favour, for example, the country with the most important consumer market. 

\section*{Acknowledgements}
This work is supported by ``Programa de Incentivo \`a Produtividade em Ensino
e Pesquisa'' from PUC-Rio.


\end{document}